# Energy Scheduling for Residential Distributed Energy Resources with Uncertainties Using Model-based Predictive Control

Anahita Moradmand, Mehrdad Dorostian, and Bahram Shafai

***Abstract*—This paper proposes a reliable energy scheduling framework for distributed energy resources (DER) of a residential area to achieve an appropriate daily electricity consumption with the maximum affordable demand response. Renewable and non-renewable energy resources are available to respond to customers' demands using different classes of methods to manage energy during the time. The optimal operation problem is a mixed-integer-linear-programming (MILP) investigated using model-based predictive control (MPC) to determine which dispatchable unit should be operated at what time and at what power level while satisfying practical constraints. Renewable energy sources (RES), particularly solar and wind energies recently have expanded their role in electric power systems. Although they are environment friendly and accessible, there are challenging issues regarding their performance such as dealing with the variability and uncertainties concerned with them. This research investigates the energy management of these systems in three complementary scenarios. The first and second scenarios are respectively suitable for a market with a constant and inconstant price. Additionally, the third scenario is proposed to consider the role of uncertainties in RES and it is designed to recompense the power shortage using non-renewable resources. The validity of methods is explored in a residential area for 24 hours and the results thoroughly demonstrate the competence of the proposed approach for decreasing the operation cost.**

## I. Introduction

THE growing electricity demand is a challenge as it not only involves the search for new sources of energy but also emphasizes on energy management. According to recent researches, up to 40% of the energy consumption takes place in both residential and non-residential buildings [1]; hence, the concept of energy scheduling caught the attention of researchers during years. In general, there are conflicting optimization goals where maximization of user comfort and minimization of energy consumption are desired. Apart from this, other additional aims for optimization such as environmental pollution have been introduced which is not satisfactory especially in the long term [2].

Through years, many approaches developed in energy scheduling. The first ideas emerged before the prevalence of renewable sources where the typical one was offline optimization. This problem is solved once for the whole period of the prediction horizon, while the optimization of the problem must be updated at every time step of the operation period (e.g., at every hour) due to the time-varying feature of the power of resources or load demands. The efficiency of offline methods may not be satisfying in serving power as they forecast the resources and demands deterministically which does not fit perfectly in practice and violates the real-world constraints. Neglecting uncertainties, inaccurate grid components modeling, and omitting operational constrains are the significant factors in deficiency of offline optimization [3-8]. Over the years, to improve the performance of these methods they were implemented with stochastic programming using Monte Carlo simulation. Although the productivity of these methods was escalated, they became computationally expensive to use [9-12].

Renewable resources are considered as the fastest-growing energy source in the United States, increasing 100 percent from 2000 to 2018 [13]. They are clean, free sources of energy, environmentally friendly, and easily accessible; moreover, they carry a huge potential in residential energy management to decrease optimization costs [14-17]. However, there are critical stability challenges regarding the employment of RES in the real-world. One of the threats arises from the inherent uncertain and random nature of these resources. This fact makes the attributes of scheduling with RES a stochastic optimization problem. Even though disregarding RES characteristic does not appear in theory, it is fruitless in reality [18-20].

The idea of optimization evolved using robust optimization to deal with uncertainties in energy scheduling [21], [22]. To do this, an uncertain set was built to describe the variability of RES. Next, the robust optimization problem was transformed into a deterministic problem by defining uncertain factors in deterministic status but the key problem lies in how to reduce a large number of scenarios. In this way, several reduction techniques like interval linear programming [23], backward and

Anahita Moradmand and Mehrdad Dorostian are with the Department of Electrical and Computer Engineering at the Northeastern University, Boston MA, USA. (email: moradmand.a, dorostian.m@northeastern.edu)

Bahram Shafai is with the Department of Electrical and Computer Engineering and the Department of Biomedical Engineering Department at the Northeastern University, Boston MA, USA. (email: shafai@ece.neu.edu)

forward scenario reduction technique [24], worst-case reduction method [25], etc. have been introduced. Compared to deterministic cases, robustness is an important factor in energy management influenced by uncertainties as it requires an accurate estimation of uncertainty bounds [20], [24], [26]. Apart from robust optimization, many stochastic and fuzzy optimizations were proposed [27-29]. Most of these approaches were not comprehensive as they generally consider load and wind power uncertainties where the solar power uncertainty was neglected. In addition to that, there are many practical constraints regarding WT and PV required to be satisfied in applicable methods. The noteworthy constraint regarding RES is that they are inoperative during a certain period of a day. Proficient optimization methods, use dynamical RES models addressing their functional constrains [30],[31].

Furthermore, the power resources are not the only effective side of distributing energy. The market economy model is a key aspect of the scheduling problem. In most researches, the price determined for customers is constant for both purchasing and selling electricity. On the other hand, there are studies distinguishing market price in two or more bids based on the time-of-use of power, namely, off-peak, mid-peak, and on-peak hours [32-35].Hourly varying prices enables cost savings by shifting substituting appliances during the time, while a constant price does not provide this flexibility. In most cases, costumers can not participate to determine market prices in every time interval.

In recent scheduling methods, to carry out challenges in energy scheduling, MPCs have been involved. These controllers can successfully incorporate in both forecasts and newly updated information to decide the future behaviors of the system and handle different kinds of operational constraints. MPC compromises a class of control algorithms that utilizes an online process model to optimize the future response of a plant. The main benefits of MPC are the explicit consideration of dynamics, available predictions of future disturbances, constraints, and conflicting optimization goals to provide the optimal control input. It is known as one of the most promising methods for dealing with forecast uncertainties in the real-time control of dynamical systems. Especially the knowledge and use of future disturbances in the optimization makes MPC such a powerful and valuable control tool in the area of building automation. In addition to that, MILP approaches were raised attention as they are capable to optimize at each decision time, based on the short-term forecasts of power generation, load demand, and electricity price [36-41].

This paper introduces a novel energy scheduling approach to overcome imperfections in previous methods. In this study, the aforementioned challenges are addressed in three different scenarios. In all the scenarios the scheduling problem is mathematically formulated and solved by MPC using MILP. Online multi-objective optimization is provided to maximize user comfort and minimize energy consumption based on 24-hour ahead forecast data. The first scenario distributes energy in a constant price market. The contribution of each renewable and non-renewable power unit is determined by forecasting future operational costs in each time scale. The second scenario also takes advantage of dynamical market prices where consumers participate in price market decisions. The third scenario is a robust optimization which prevents system failure due to energy shortage in renewable resources by applying more capacity of non-renewable sources. To that, a dynamical model of PV and WT is considered to obtain uncertainty boundaries. The differences in the scenarios are advantageous in their suitable applications.

The overview of the scheduling schematic of this paper is shown in Figure 1. The power side, depicted on top, consists of renewable resources and non-renewable resources. The RESs are PVs and WTs which are prioritized to battery energy storage systems (BESSs) and distributed generators (DGs) as non-renewable resources. On the consumer side, there are electrical vehicles (EV), industrial consumers, and controllable loads (CL) where CLs mostly involve thermostatically controlled loads (TCL), air conditioners (AC), heating, ventilation, and air conditioning (HVAC). The goal is to manage the population and price of DERs to achieve maximum demand response. All of them can share their own desired price to grid and base on resilient and economic features decide about their shared or consumed powers with or from the main grids. Based on the interference of uncertainty and dynamic of the market price in the optimization problem, different scenarios are proposed. In all scenarios, the information and power provided by both sources and sinks are gathered in a SCADA block and the MPC and its criteria in each scenario decide on energy distribution. Also, the Management block defines the operational costs of DERs based on the data given by the SCADA block. In each time interval, the MPC determines the participation of power resources using the optimization cost defined in each scenario.

The rest of the paper is structured as follows. Section 2 explains the problem. In section 3, the optimization scenarios are explained and the validated simulation result are described in section 4.

## II. Problem Formulation

Suppose a residential microgrid, shown in Figure 1, which is equipped with renewable (PV and WT) and non-renewable resources (BESS and DG). Consumer loads mainly are TCLs, HVACs, ACs on the domestic side, and factories on the industrial side. A market-based solution to the energy scheduling problem of these residential buildings microgrid systems which can serve a peak load capacity of up to 6MW is considered here. MILP energy scheduling is investigated using MPC in 3 different scenarios. It aims at minimizing the overall operation cost by controlling both supply and demand-side units. The mathematical models of the microgrid units, power balance constraints, and objective function are presented in this section.

In this research, renewable resources are prioritized to take action and the contribution of other resources depends on the availability of renewable resources. The total optimization cost $J$ is a combination of three sub cost functions. The optimization problem is solved in a time scale of $N_k$ and in each step, if the RES offers sufficient power, the other resources decide whether to work in no-load status or turn off based on the minimized cost in $J_2$. Apparently, if RES is unable to satisfy the consumers, the system profits from DGs and BESSs and their on-state performance costs are considered in $J_1$. Also, the market price has dynamics in two scenarios which helps in maximizing the affordable power in each step. To do this, the DERs state of charge is considered to be maximized in $J_3$. The cost functions are mathematically modeled in equations (1) to (4)

$$J = J_1 + J_2 + J_3 \tag{1}$$

$$J_1 = \sum_{k=1}^{N_k} \sum_{i=1}^{N_{DG}} C_{i,k}^{DG} P_{i,k}^{DG} + \sum_{k=1}^{N_k} \sum_{i=1}^{N_{BESS}} C_{i,k}^{BESS} P_{i,k}^{BESS} \tag{2}$$

$$J_2 = \sum_{k=1}^{N_k} \sum_{i=1}^{N_{DG}} (C_{i,k}^{Start} ST_{i,k} + C_{i,k}^{No-load} ON_{i,k}) \tag{3}$$

$$J_3 = \sum_{j=1}^{N_{DER}} SOC_{i,k} \tag{4}$$

where $J$ is total cost function, $J_1$ is costs of non-renewable power supplies, $J_2$ is costs of starting and no-load working status of DGs and BESSs, and $J_3$ is DERs total state of charges. The state dynamics of DERs (TCLs, electric vehicles, and batteries) in (4) are expressed in (5) using a generalized battery model presented in [42-44]. Each DER's State of Charge (SOC) varies between its minimum and maximum values and in this research it is normalized between 0 and 1. Furthermore, $p_{i,k}$ is the $i$th DER's price at time $k$, which normally decreases as $SOC_{i,k+1}$ increases according to (6). The feeder, to which DERs are connected, sets a market price to afford a specific demand in different iterations. In the first scenario, this $p_k^{clear}$ is assumed to be constant while it varies in the second and the third scenarios.

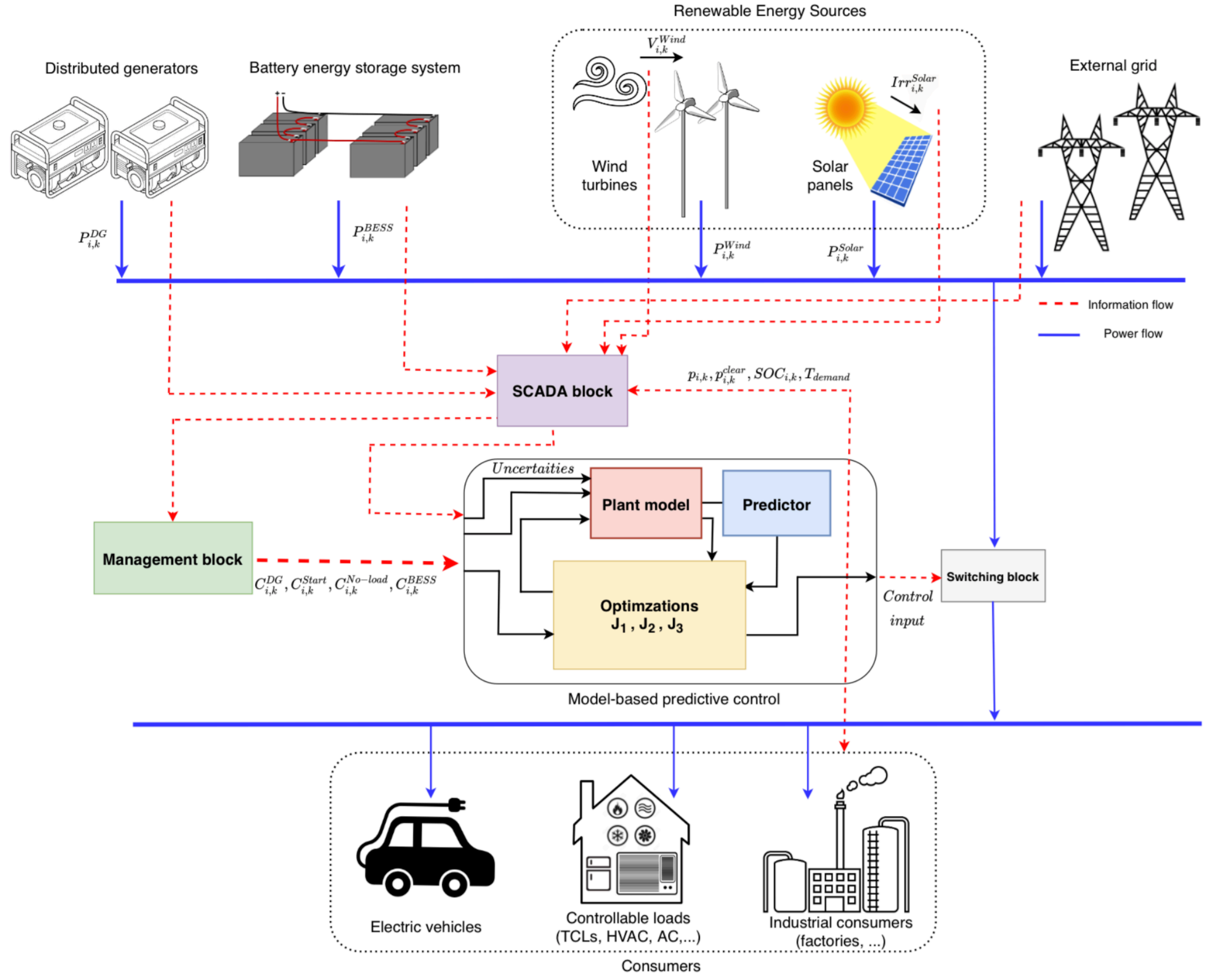

Figure 1. Schematic model

$$SOC_{i,k+1} = a_i SOC_{i,k} + \gamma_i m_{i,k} v_{i,k} \tag{5}$$

$$p_{i,k} = p_{max} - \beta_i SOC_{i,k} \tag{6}$$

$$v_{i,k} = \begin{cases} 0 & if \quad p_{i,k} < p_k^{clear} \\ 1 & if \quad p_{i,k} \geq p_k^{clear} \end{cases} \tag{7}$$

$$m_{i,k+1} = \begin{cases} 0 & if \quad SOC_{i,k} \geq SOC_{i,max} \\ 1 & if \quad SOC_{i,k} < SOC_{i,set} \\ m_{i,k} & if \quad O.W. \end{cases} \quad (8)$$

Also, there are practical constrains to equations (5) to (8) as defined in (9) to (12).

$$SOC_{i,set} \leq SOC_{i,k} \leq SOC_{i,max} \quad (9)$$
$$\alpha_{i,min} \leq \alpha_i \leq \alpha_{i,max} \quad (10)$$
$$\beta_{i,min} \leq \beta_i \leq \beta_{i,max} \quad (11)$$
$$\gamma_{i,min} \leq \gamma_i \leq \gamma_{i,max} \quad (12)$$

Table 1 represents cost functions and optimization variables.

## III. Energy Scheduling Scenario

The MILP energy scheduling is performed using MPC in the three following scenarios. In each case, the power of RES side in $N_k$ steps are estimated using PV and WT models. This estimation varies among scenarios provided that RESs act deterministically or non-deterministically. Also, MPC determines DERs' SOC optimizing $J_3$ using the only varying parameter $p^{clear}$ which is considered to be constant or inconsistent. Taking advantage of SOC and RES power, MPC figures out the contribution of non-renewable sources using $J_1$ and $J_2$ as described in the following. In time step $t_1$ the optimization problem is solved in the first horizon to $t_{f1}$ and the MPC decides over BESSs and DGs participation at this step by maximizing $J_3$ and minimizing $J_1$ and $J_2$ in this control horizon. Then, we proceed to $t_1 + 1$ and the optimization horizon would be $t_1 + 1$ to $t_{f1}$. The MPC takes place in each $N_k$ time scale as shown in Figure 2 where at each time, the MPC calculates optimization cost $J$.

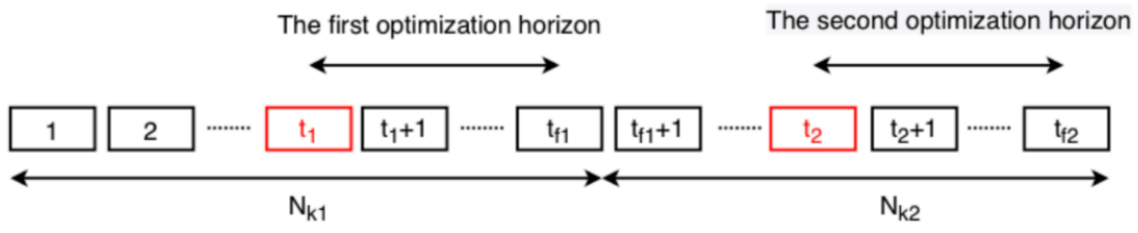

Figure 2. Optimization horizon

To solve a feasible optimization problem, the power provided in DGs and BESSs is classified into bids and they are not expected to cover a power between every two consecutive bids as shown in Figure 3. Also, when DGs are workless at a time step they will be turned off or continue in the no-load state; hence, the $ST$ and $NO$ in $J_2$ represent integer parameters and the MPC is supposed to deal with a MILP optimization problem. The scheduling scenarios are explained in the following.

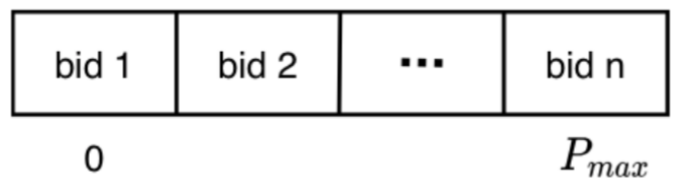

Figure 3. DG and BESS bids

### A. *Scenario 1: Constant clearing price and deterministic RES sources*

In this case, the power provided in the RES side is assumed to be deterministic. Additionally, the market price is constant and the consumers do not participate in determining it; hence, the optimization in $J_3$ is disregarded as there is no control over the market price to maximize the level of SOC.

- *Deterministic photovoltaic panels and wind turbines*

The deterministic behavior of RES sources are described in

$$P^{Wind}_{i,k+1} = \begin{cases} P^{Wind}_{i,max} & if \quad v^{Wind}_{i,max} \leq v^{Wind}_{i,k+1} \\ P^{wind}_{i,k} & if \quad v^{Wind}_{i,min} \leq v^{Wind}_{i,k+1} \leq v^{Wind}_{i,max} \\ 0 & if \quad v^{Wind}_{i,k+1} < v^{Wind}_{i,min} \end{cases} \quad (13)$$

$$P^{Solar}_{i,k+1} = \begin{cases} P^{Solar}_{i,max} & if \quad Irr^{Solar}_{i,max} \leq Irr^{Solar}_{i,k+1} \\ P^{Solar}_{i,k} & if \quad Irr^{Solar}_{i,min} \leq Irr^{Solar}_{i,k+1} \leq Irr^{Solar}_{i,max} \\ 0 & if \quad Irr^{Solar}_{i,k+1} < Irr^{Solar}_{i,min} \end{cases} \quad (14)$$

Also, the output power should not exceed their corresponding design limit at each time step

$$P^{Solar}_{i,k} \leq P^{Solar}_{i,max} \quad (15)$$
$$P^{Wind}_{i,k} \leq P^{Wind}_{i,max} \quad (16)$$

As demonstrated, the PVs' and WTs' power is directly influenced by irradiance and wind speed, respectively which are collected in time scale $N_k$. Determining the RES power in (13) and (14), the optimization specifically takes place through $J_1$ and $J_2$ in (2) and (3). Although this scenario does not consider dynamics in market price and uncertainty in RES, it effectively decreases the amount of calculation.

### B. *Scenario 2: Varying clearing price and deterministic RES sources*

The second scenario participates in an active pricing market which is able to change its energy consumption based on the cleared market price and determine the price it is willing to pay for electricity. Also, dynamic market pricing allows to create an interactive system which can limit demand at key times on a distribution system. The price market determination is described in the following.

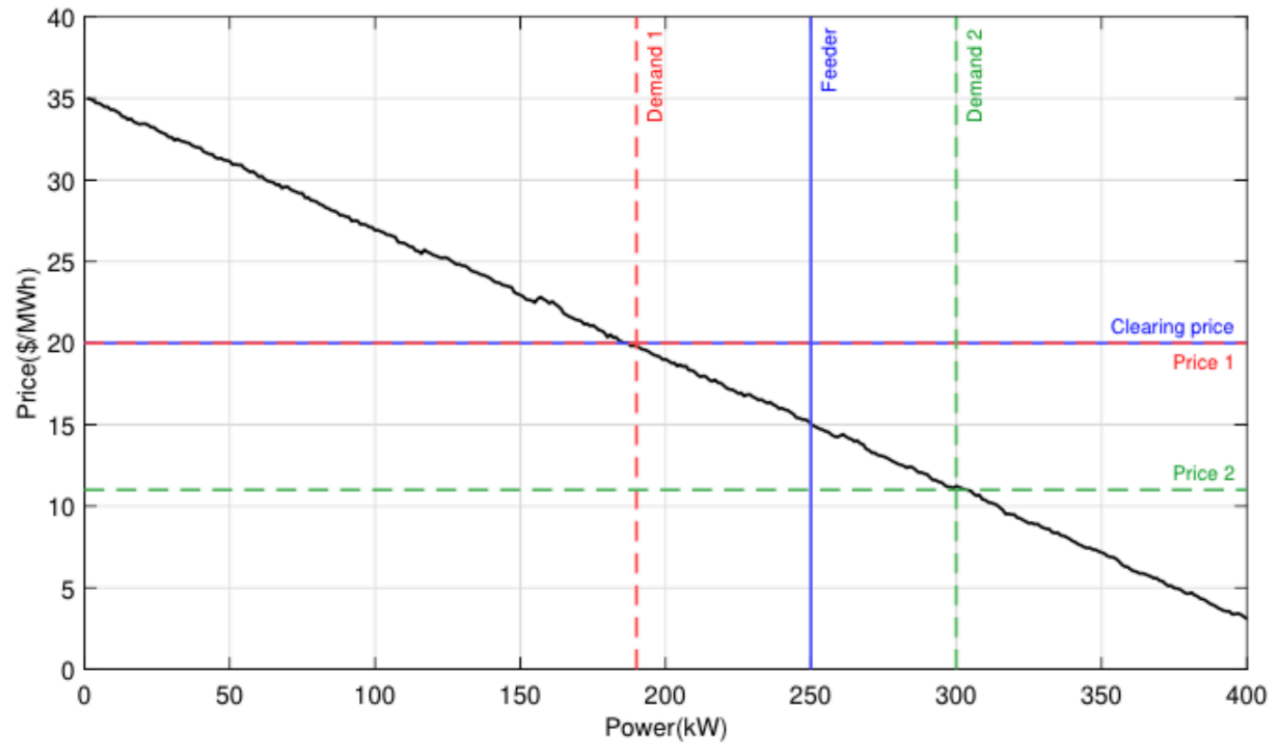

Figure 4. Market clearing price

| | | | |
|---|---|---|---|
| $k$ | Time interval index | $N_k$ | Number of time intervals |
| $J$ | Total cost function | $J_1$ | Total cost of power supplies |
| $J_2$ | Total cost of working under no-load or start state | $J_3$ | Total power demand |
| $N_{DG}$ | Number of DGs | $N_{BESS}$ | Number of BESSs |
| $P_{i,k}^{DG}$ | Power generated by *i*th DG unit at time interval *k* | $P_{i,k}^{BESS}$ | Power generated by *i*th BESS unit at time interval *k* |
| $P_{i,max}^{DG}$ | Maximum power generated by *i*th DG unit | $P_{i,max}^{BESS}$ | Maximum power generated by *i*th BESS unit |
| $C_{i,k}^{DG}$ | *i*th DG power supply production cost at time interval *k* | $C_{i,k}^{DG}$ | *i*th BESS power supply production cost at time interval *k* |
| $ST_{i,k}$ | Start status of the *i*th DG at time interval *k* (ON/OFF) | $ON_{i,k}$ | No-load status of the *i*th DG at time interval *k* (ON/OFF) |
| $C_{i,k}^{No-load}$ | No-Load cost of the *i*th DG at time interval *k* | $C_{i,k}^{Start}$ | Start cost of the *i*th DG at time interval *k* |
| $N_{DER}$ | Number of DERs | $SOC_{i,k}$ | The *i*th load state of charge at interval *k* |
| $a_i$ | Energy dissipation rate of *i*th DER | $\gamma_i$ | Charging efficiency of *i*th DER |
| $m_{i,k}$ | lock-out state of *i*th DER at interval *k* | $v_{i,k}$ | Power ON/OFF decision of *i*th DER at time interval *k* |
| $p_{i,k}$ | The *i*th DER market price at interval *k* | $\beta_i$ | The *i*th DER price to SOC ratio |
| $p_k^{clear}$ | Market clear price at interval *k* | $p_{max}$ | Maximum consumer price |
| $SOC_{i,set}$ | SOC of *i*th DER set value | $SOC_{i,max}$ | SOC of *i*th DER maximum value |
| $a_{i,min}$ | Minimum energy dissipation rate of *i*th DER | $a_{i,max}$ | Maximum energy dissipation rate of *i*th DER |
| $\gamma_{i,min}$ | Minimum charging efficiency of *i*th DER | $\gamma_{i,max}$ | Maximum charging efficiency of *i*th DER |
| $\beta_{i,min}$ | The *i*th DER minimum price to SOC ratio | $\beta_{i,max}$ | The *i*th DER maximum price to SOC ratio |
| $N_{Solar}$ | Number of solar units | $N_{Wind}$ | Number of wind units |
| $P_{i,k}^{Solar}$ | Power generated by *i*th solar unit at time interval *k* | $P_{i,k}^{Wind}$ | Power generated by *i*th wind unit at time interval *k* |
| $P_{i,max}^{Solar}$ | Maximum power generated by *i*th solar unit | $P_{i,max}^{Wind}$ | Maximum power generated by *i*th wind unit |
| $C_{i,k}^{Solar}$ | *i*th solar power supply production cost at time interval *k* | $C_{i,k}^{Wind}$ | *i*th wind power supply production cost at time interval *k* |
| $Irr_{i,k}^{Solar}$ | Irradiance of *i*th solar unit at time interval *k* | $v_{i,k}^{Wind}$ | Wind speed *i*th wind unit |
| $Irr_{i,min}^{Solar}$ | Minimum irradiance *i*th solar unit | $Irr_{i,max}^{Solar}$ | Maximum irradiance of *i*th solar unit |
| $v_{i,min}^{Wind}$ | Minimum wind speed *i*th wind unit | $v_{i,max}^{Wind}$ | Maximum wind speed *i*th wind unit |

Table 1. Optimization parameters

- *Market clearing price*

Let $p^{base}$ be the base price forecast at the *k*th time period. All the proposed prices are gathered to build a price versus demand function. Using the price to demand diagram and the base price information $p^{base}$, the corresponding demand in this specific price is determined. If demand in $p^{base}$ is less than feeder's demand (*Demand 1* in Figure 4), then the market price is the same as $p^{base}$. Otherwise, if demand in $p^{base}$ exceeds feeder's demand (*Demand 2* in Figure 4), a $p_k^{clear}$ is set.

In this scenario the market price is classified into bids. In addition to the optimization stated in previous scenario, $J_3$ takes action and MPC assigns a market bid which can maximize the overall affordable demand in $N_k$ regarding power resources capacity.

### C. Scenario 3: Varying clearing price and uncertain RES sources

In this scenario, the dynamic market price and RES uncertainty both are considered. Uncertainty in irradiance and wind measurements is in part related to the instruments, and in part to the measurement practices. The robust strategies in power resources show their significant role when the provided power is much less than the predicted amount due to uncertainties. In the following the influence of uncertainties on PVs and WTs performance is illustrated.

- *Photovoltaic panels dynamics*

The dynamics of PV panels depend on every cell it contains [45]. The equivalent circuit of a simplified single-diode model used in PV cells is shown in Figure 5.

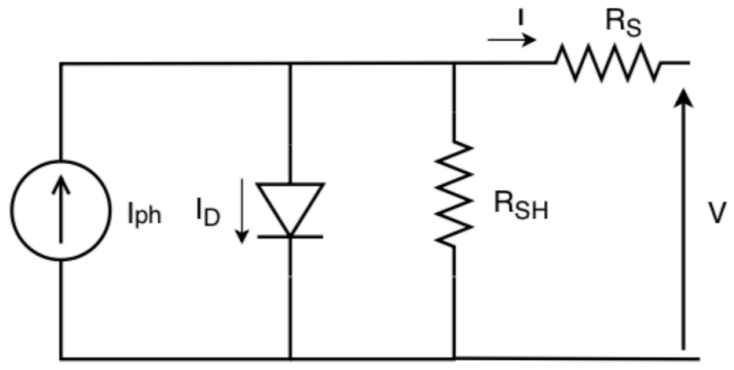

Figure 5. PV cell circuit

$$I_{ph}(G_a,T) = I_{scs}\frac{G_a}{G_{as}}[1 + \Delta I_{sc}(T - T_s)] \tag{17}$$

where $I_{ph}$, $G_a$, and $T$ are photo current, irrandiance, and temperature, respectively. Also, $I_{scs}$ is short circuit current on standard test condition, $G_a$ is standard irradiance, $\Delta I_{sc}$ is parameter coefficient on short circuit current, and $T_s$ is standard temperature. Considering equations $P = VI$ and (17), the PVs' power modification due to uncertainty in irradiance can be investigated.

- *Wind turbines dynamics*

The power generated by a WT can be modeled as

$$P = \frac{1}{2}\rho A v^3 \quad (18)$$

where $P$, $\rho$, $A$ and $A$ are power ($W$), density ($kg/m^3$), swept area, and wind speed ($m/s$). Equation (18) is derived using the kinetic energy of a mass in motions as demonstrated in [46]. WT can not convert more than 59.3% of the kinetic energy of the wind, hence; a power coefficient ($C_P$) is required to be added to (18)

$$P = \frac{1}{2}\rho A v^3 C_p \quad (19)$$

Given the wind speed and irradiance uncertainty bound, in this case the robust optimization considers the maximum power loss in RES to be addressed by non-renewable resources in equations (1) to (4).

## IV. Simulation Results and Comparison

The case study is a distributed system feeding a residential load of 1000 customers. The first power resources are PVs and WTs. If the costumers' demand is not fully responded by RES, the other resources (DGs and BESSs) will be employed. The energy scheduling of the system for a 24-hour time scale is studied here and the energy management from power resources to customers is presented. The first resource to respond to the estimated demand is RES, where if it is found insufficient the customers are managed using BESSs and DGs. The involvement of sources is decided regarding their costs as discussed in section 3.

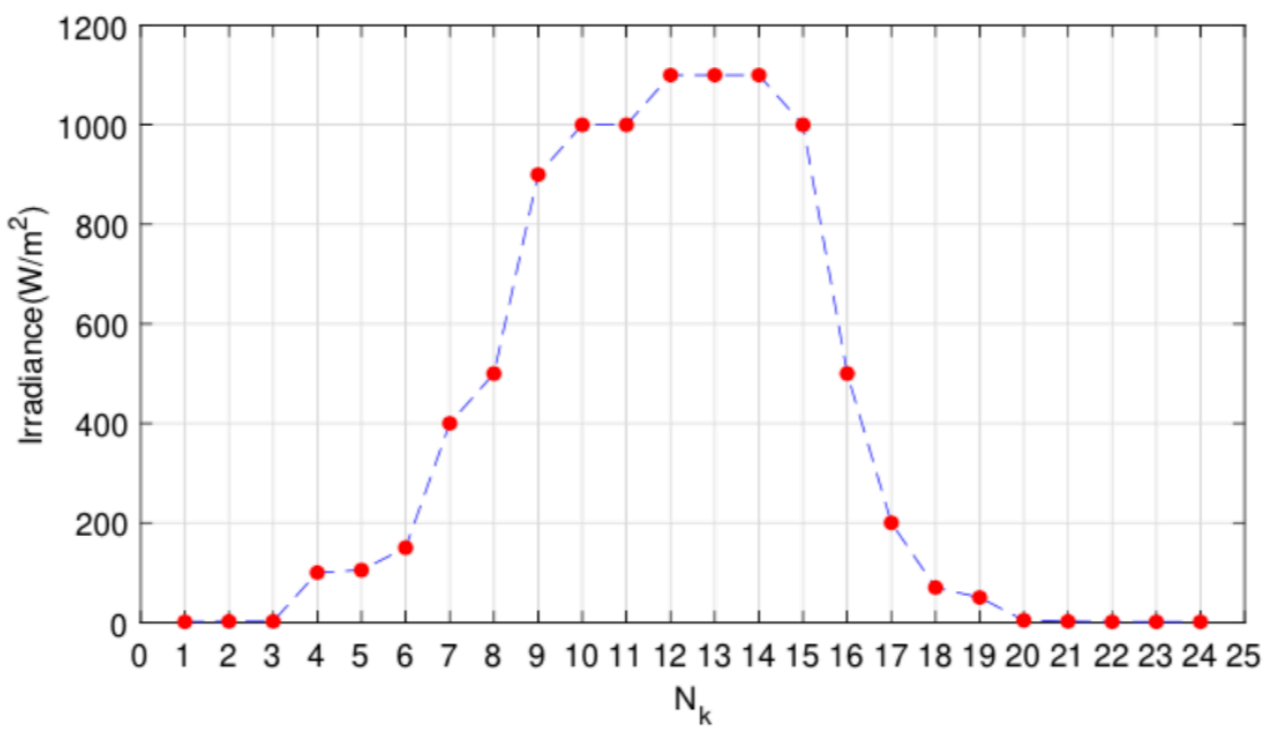

Figure 6. Irradiance in $\mathrm{N_k} = 24$

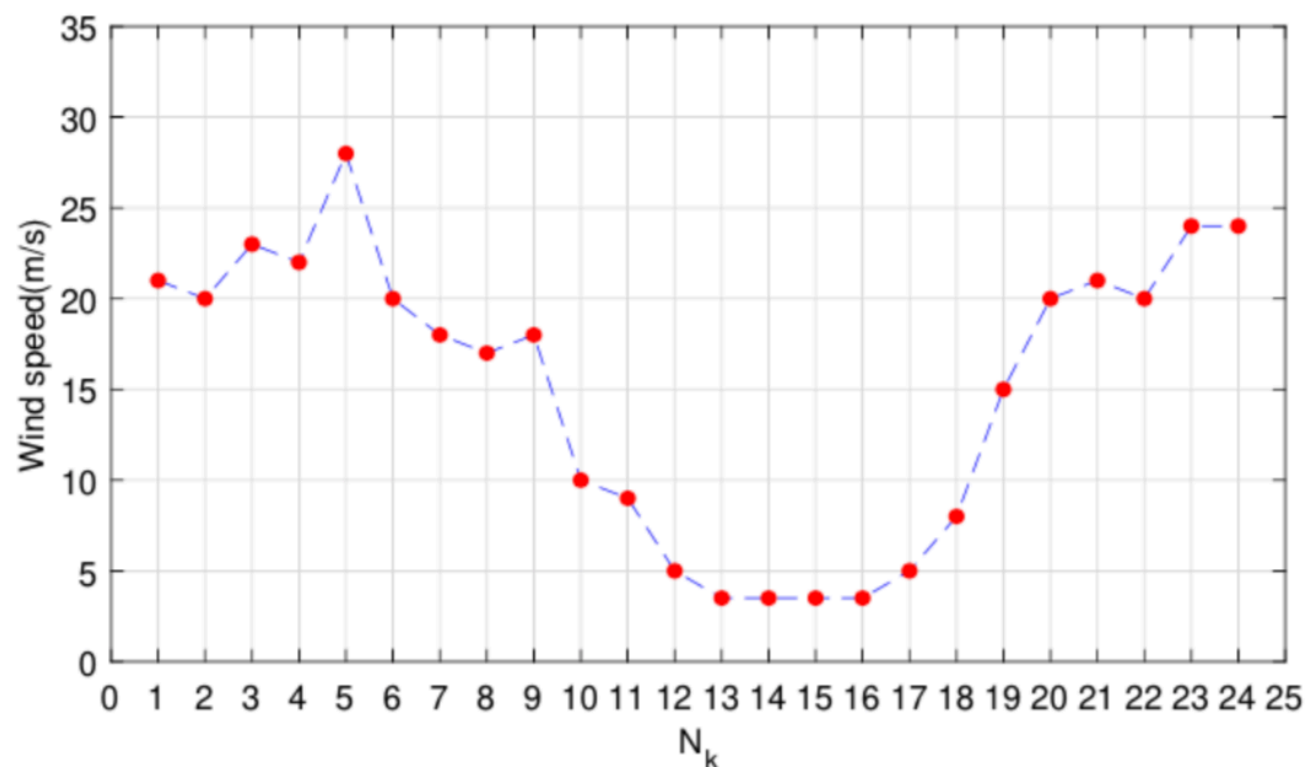

Figure 7. Wind speed in $\mathrm{N_k} = 24$

### 1) *Renewable energy sources*

In RES side, we have two 2 $MW/hour$ wind turbines and a 3 $MW$ photovoltaic resource where they can support a maximum of 6 $MW$ power in total. To estimate the performance of RES, the values of irradiance and wind speed are collected as shown in Figures 6 and 7. Obviously, the irradiance is poor at night and reaches its peak around noon as demonstrated. In Figure 7, the wind speed in the 24-hour cycle is shown which is typically higher during night hours in this investigation. The RES power shown in Figure 8 is calculated using (13) and (14) with parameters in Table 2.

| parameter | value | parameter | value |
|---|---|---|---|
| $v_{i,max}^{Wind}$ | $25m/s$ | $Irr_{i,max}^{Solar}$ | $1050W/m^2$ |
| $v_{i,min}^{Wind}$ | $3.5m/s$ | $Irr_{i,min}^{Solar}$ | $100W/m^2$ |
| $P_{i,max}^{Wind}$ | $3KW$ | $P_{i,max}^{Solar}$ | $0.1KW$ |

Table 2. RES parameters' values

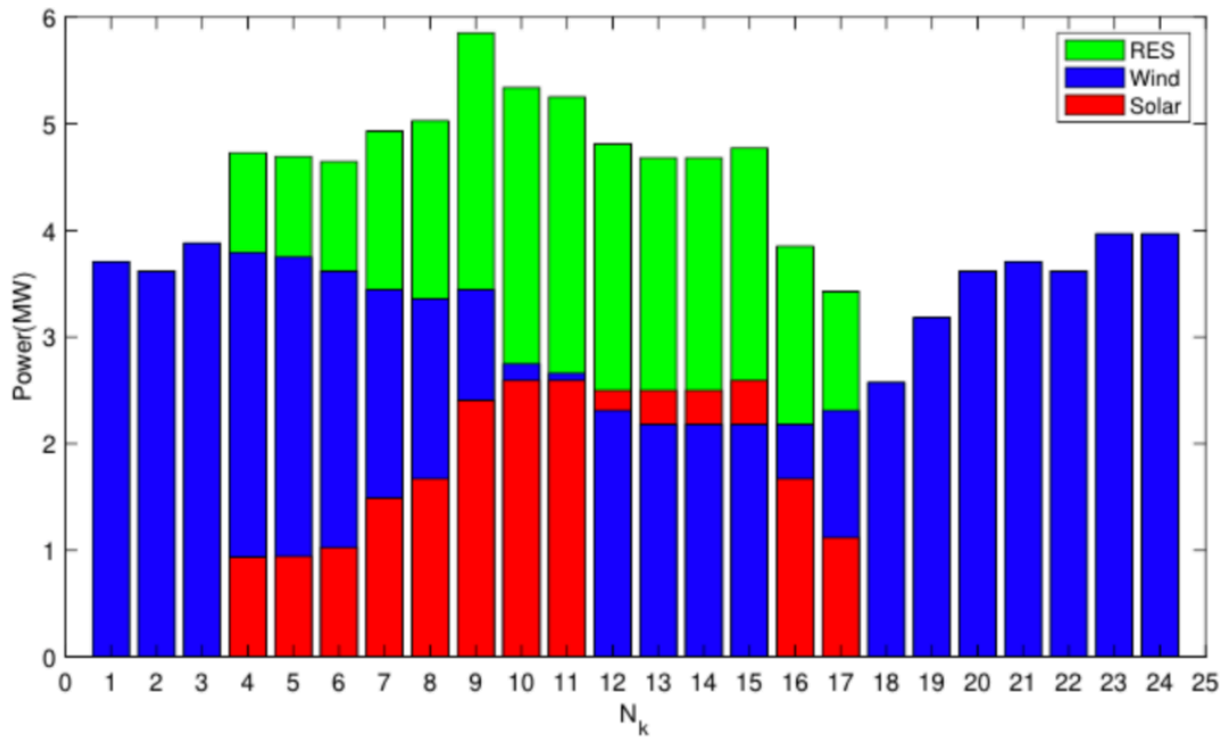

Figure 8. RES power in $\mathrm{N_k} = 24$

The RES is unable to satisfy expectations due to uncertainty in wind and irradiance in the third scenario. To consider uncertainty boundaries, the models in previous section are used. Specifying the values in Table 3, the *P-V* plots are generated as shown in Figure 9. Hence, the available power in the PV side determined using the irradiance in $N_k$ shown in Figure 6 and the plots in Figure 9. Thus, we are able to estimate the RES power influenced by uncertainty as shown in Figure 10.

| parameter | value | parameter | value |
|---|---|---|---|
| $I_{scs}$ | $7.84A$ | $G_{as}$ | $1000W/m^2$ |
| $\Delta I_{sc}$ | $0.102A$ | $T_s$ | $100W/m^2$ |
| $R_s$ | $0.393\Omega$ | $R_{sh}$ | $0.1KW$ |
| $\rho$ | $1.23kg/m^3$ | $C_p$ | 0.4 |
| $A$ | $8495m^2$ | | |

Table 3. PV and WT parameters' values

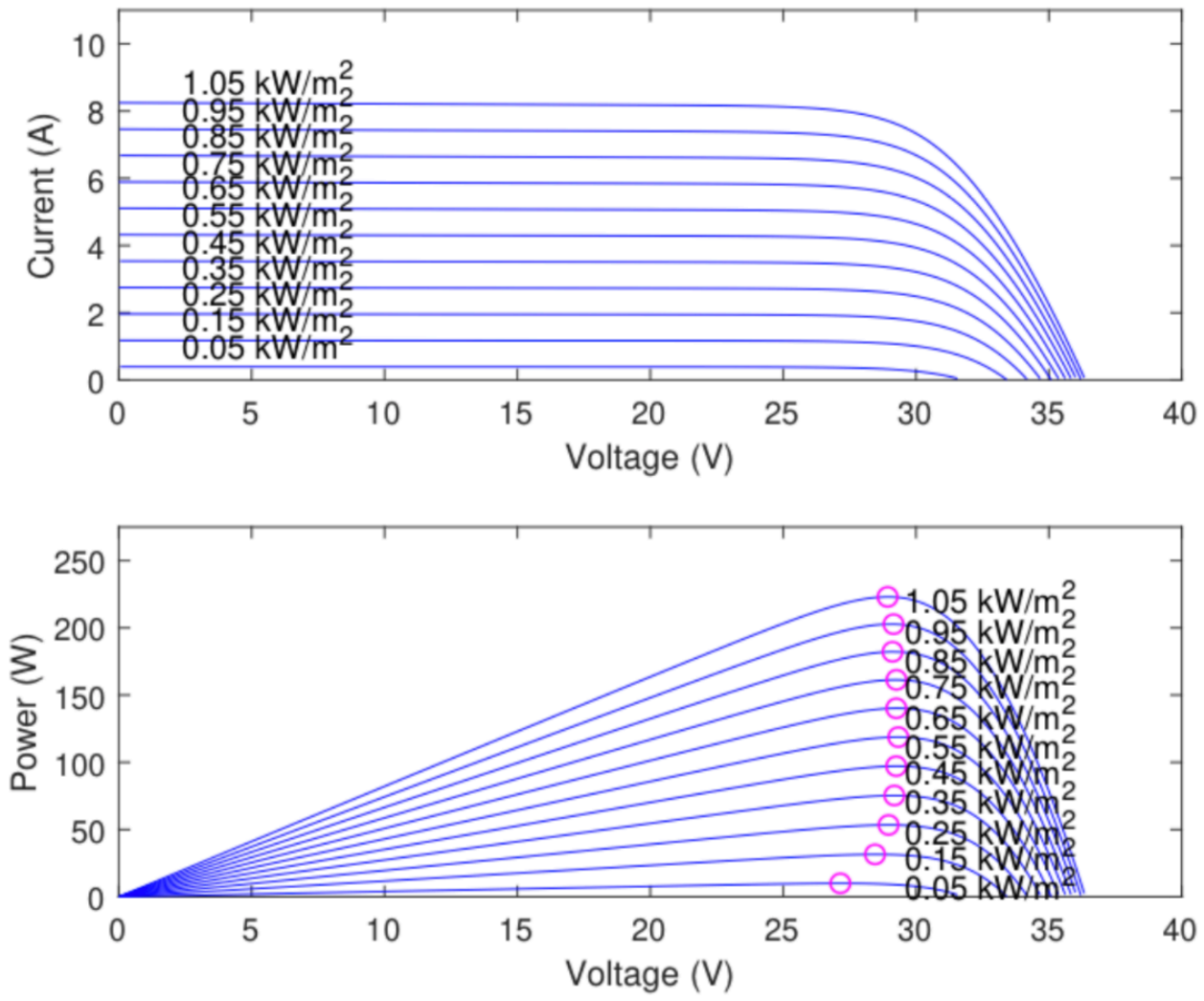

Figure 9. *P*-*V* curves

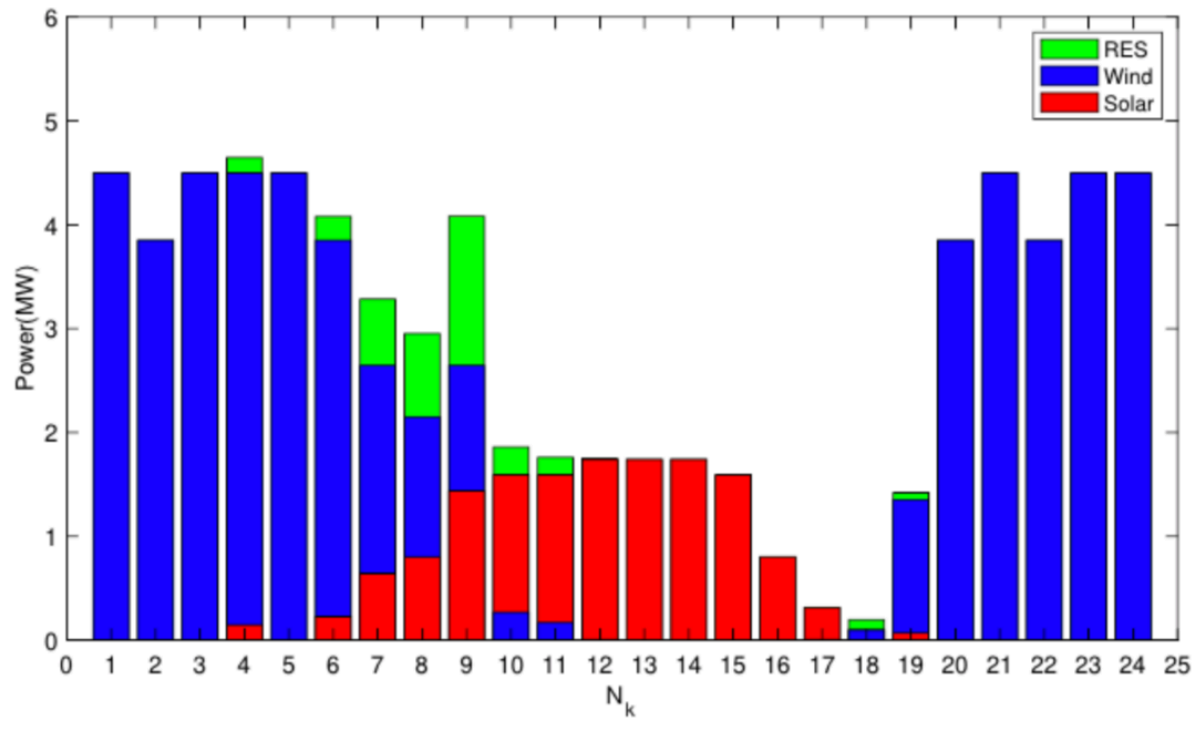

Figure 10. RES power in $N_k = 24$

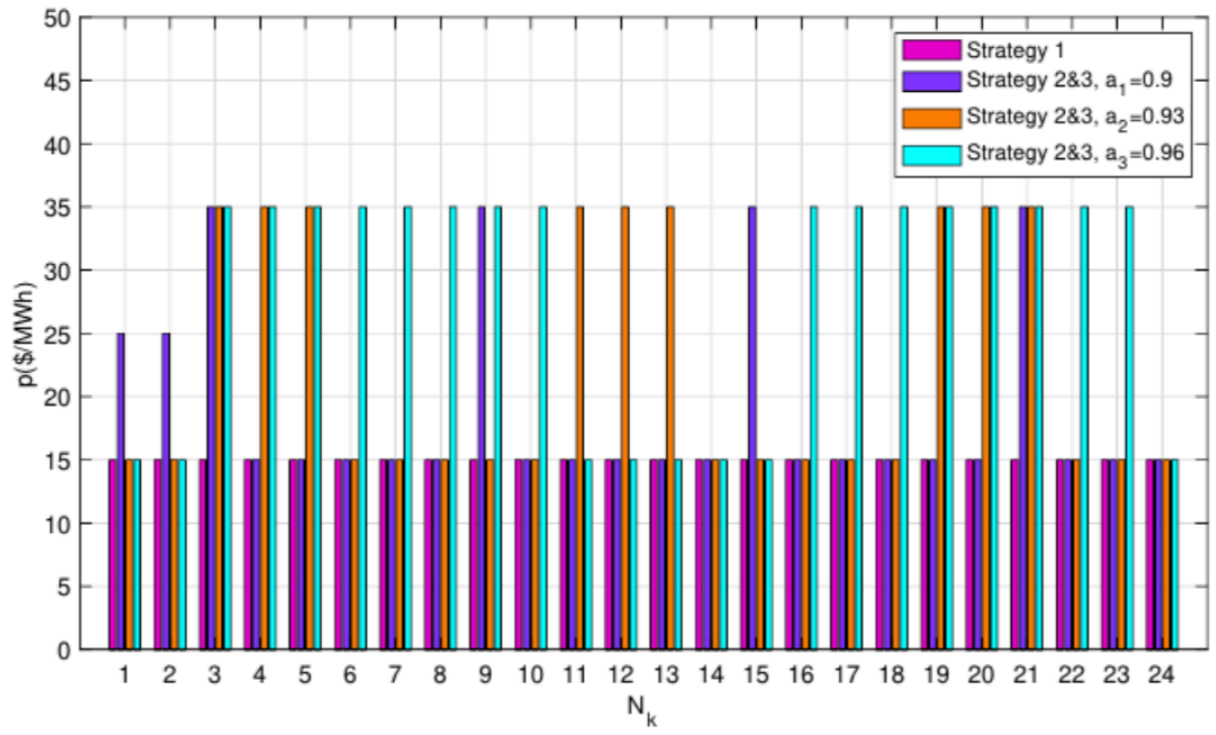

Figure 11. $p^{clear}$ in $N_k = 24$

## 2) *Market Price*

The only optimizing factor in SOC values, according to (5), is market price ($p^{clear}$). In case 1, $p^{clear}$ is assumed to be constant in $15\$/MWh$. In cases 2 and 3, the market price varies dynamically among $15\$/MWh$, $25\$/MWh$, and $35\$/MWh$. The choice of $p^{clear}$ in each $N_k$ depends on the maximum affordable demand according to $J_3$ in (4). The values of $p^{clear}$ in 24-hour cycle is demonstrated in Figure 11.

## 3) *State of charge*

The energy state of DERs, such as TCLs, electric vehicles, and batteries are modeled in (5). In this research, the DERs are classified into 3 groups with different values of energy dissipation rate ($a$) as illustrated in Table 4.

| parameter | value | parameter | value |
|---|---|---|---|
| $a_1$ | 0.9 | $a_2$ | 0.93 |
| $a_3$ | 0.96 | $\gamma$ | 1 |
| $\beta_i$ | 40 | $p_{max}$ | $30\$/MWh$ |
| $a_{min}$ | 0 | $a_{max}$ | 1 |
| $\beta_{min}$ | 0 | $\beta_{miax}$ | 1 |

Table 4. SOC parameters' values

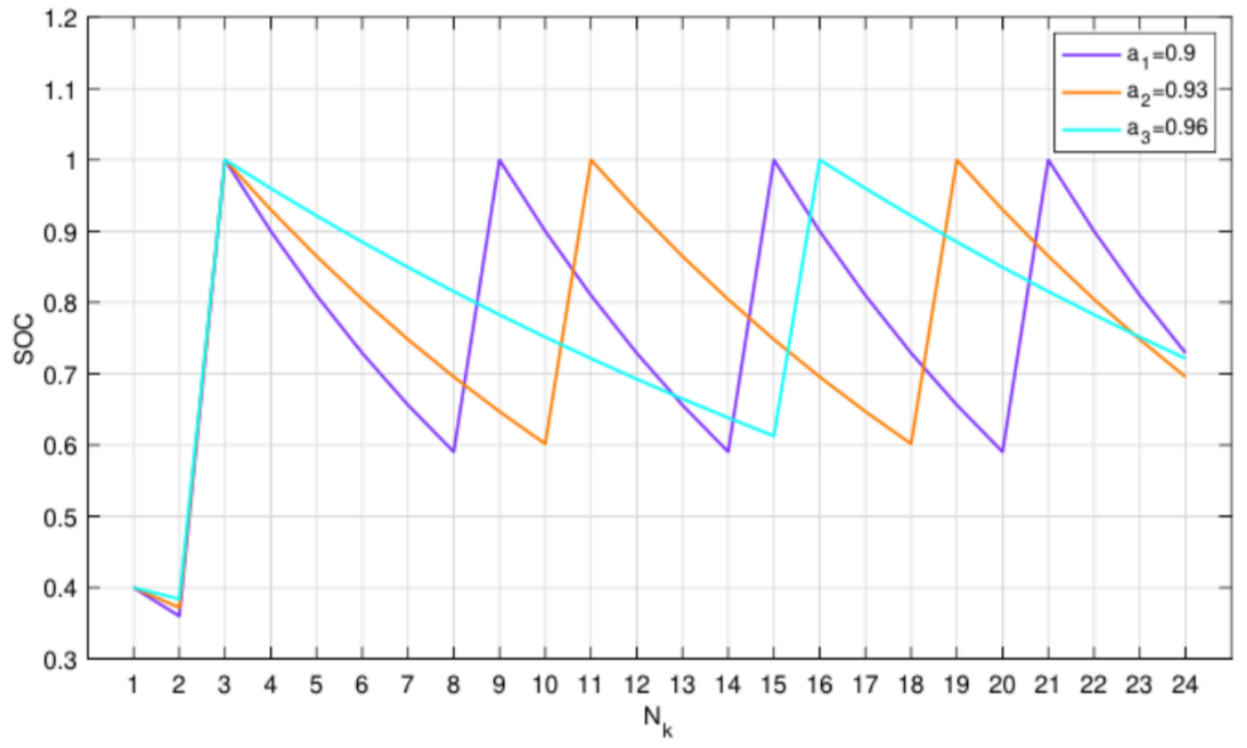

Figure 12. SOC in $N_k = 24$ while $p^{clear}$ is constant

If the DERs are always connected to power resources their SOC will stick to 1. To use the energy stored in DERs, they will be recharged when their power fall below the set value 0.7 to be charged to maximum.

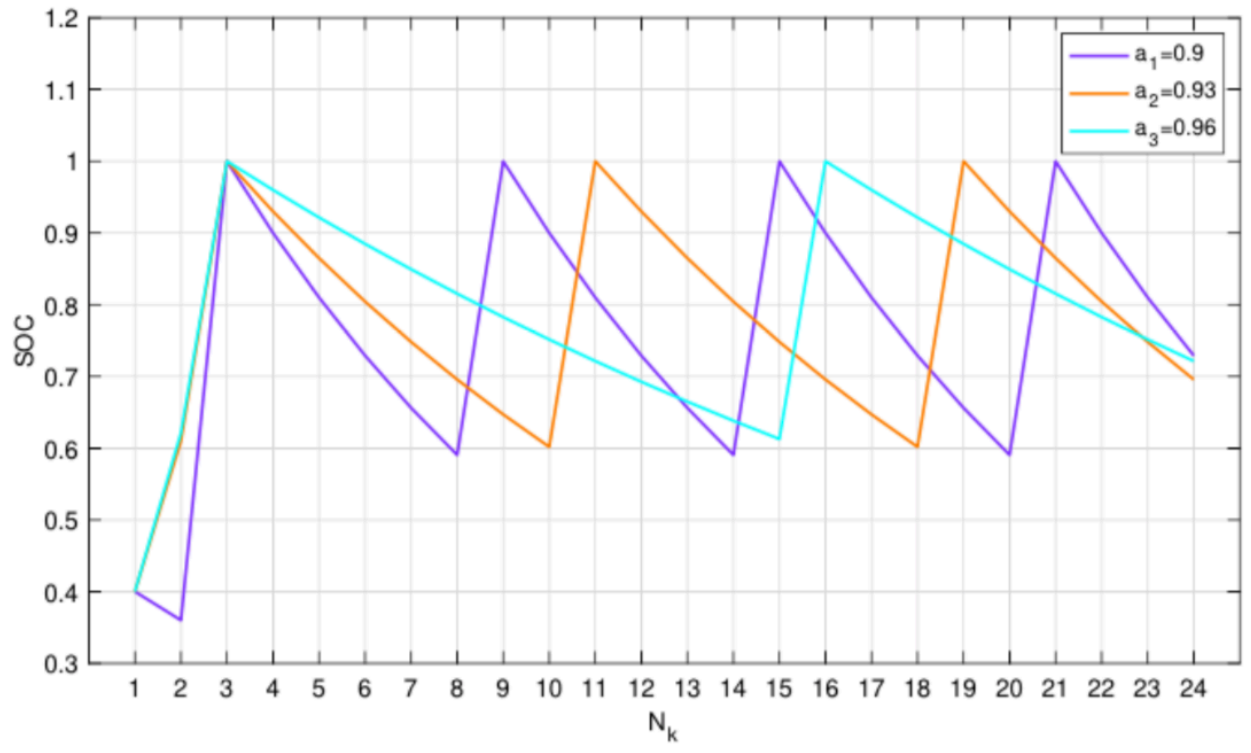

Figure 13. SOC in $N_k = 24$ while $p^{clear}$ is not constant

Accordingly, the price costumers pay in (6) with or without different values of market price are shown in Figures 14 and 15. Considering equations (5) and (6) the price logically increases in the opposite direction of SOC as illustrated in Figures 12 to 15.

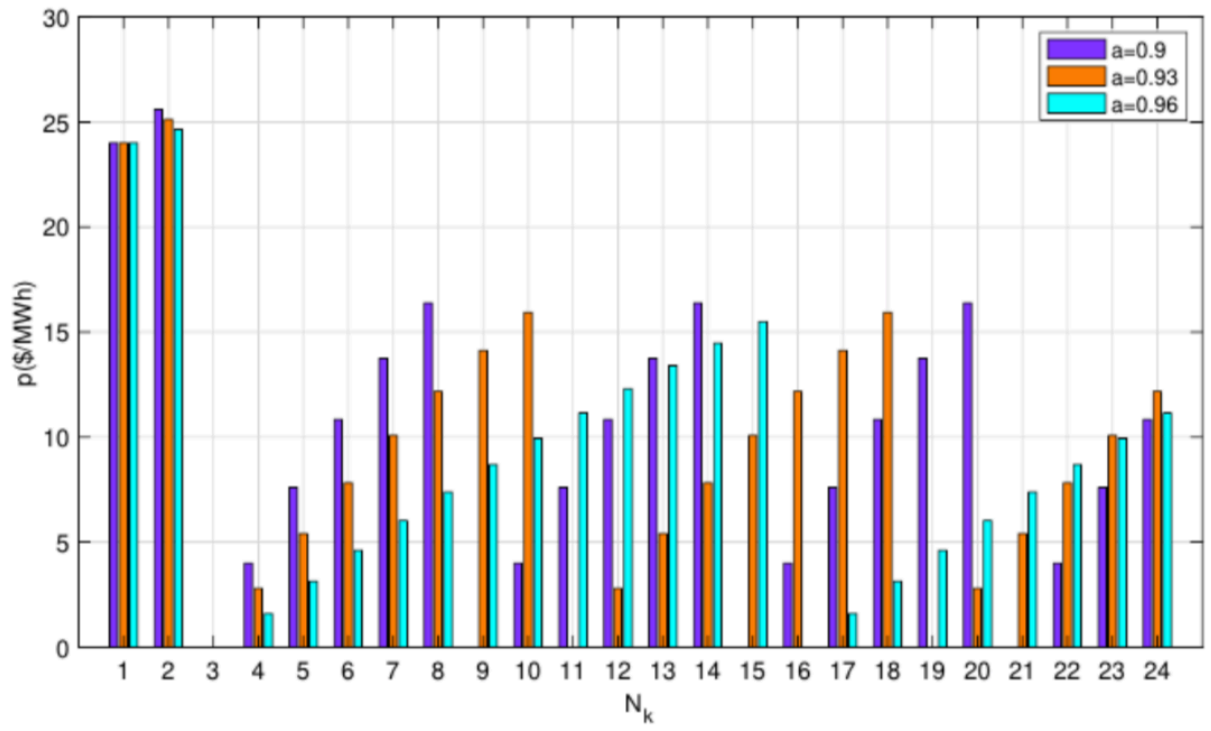

Figure 14. Price in $N_k = 24$ while $p^{clear}$ is constant

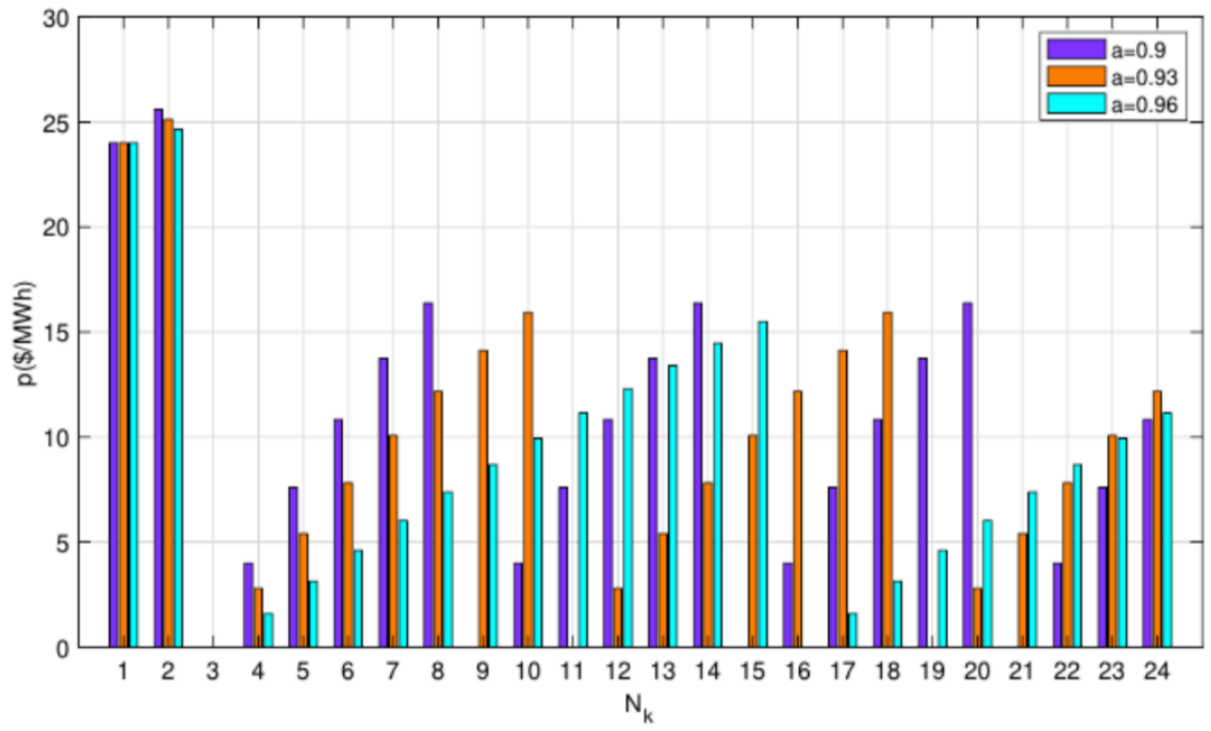

Figure 15. Price in $N_k = 24$ while $p^{clear}$ is not constant

*4) Cost function*

In energy management, the aim is to maximize the demand response and minimize optimization costs. In (1), $J_3$ maximizes the affordable demand and $J_1$ and $J_2$ minimize costs in DGs and BESSs. In $J_3$, the demand is maximized based on the DERs state of charge where the only control parameter is market price. As the market price is constant in case 1, only $J_1$ and $J_2$ are involved in cost function; hence, the values of cost functions in cases 2 are greater than 1 as demonstrated in cumulative Figures 16 and 17. In case 3, due to uncertainty in renewable resources, the demand is mainly responded by non-renewable sources and it provides more flexibility in $J_1$ and $J_2$; hence, the cost function costs decrease in the third scenario as depicted in Figure 18. In these figures the Table 5 contains the values specified to parameters in equations (1) to (4).

| parameter | value | parameter | value |
|---|---|---|---|
| $C_{i,k}^{DG}$ | 1 | $C_{i,k}^{BESS}$ | 0.1 |
| $C_{i,k}^{Start}$ | 2 | $C_{i,k}^{No-load}$ | 1 |
| $N_{DER}$ | 1000 | $N_{DG}$ | 10 |
| $N_{BESS}$ | 50 | | |

Table 5. DG and BESS parameters' values

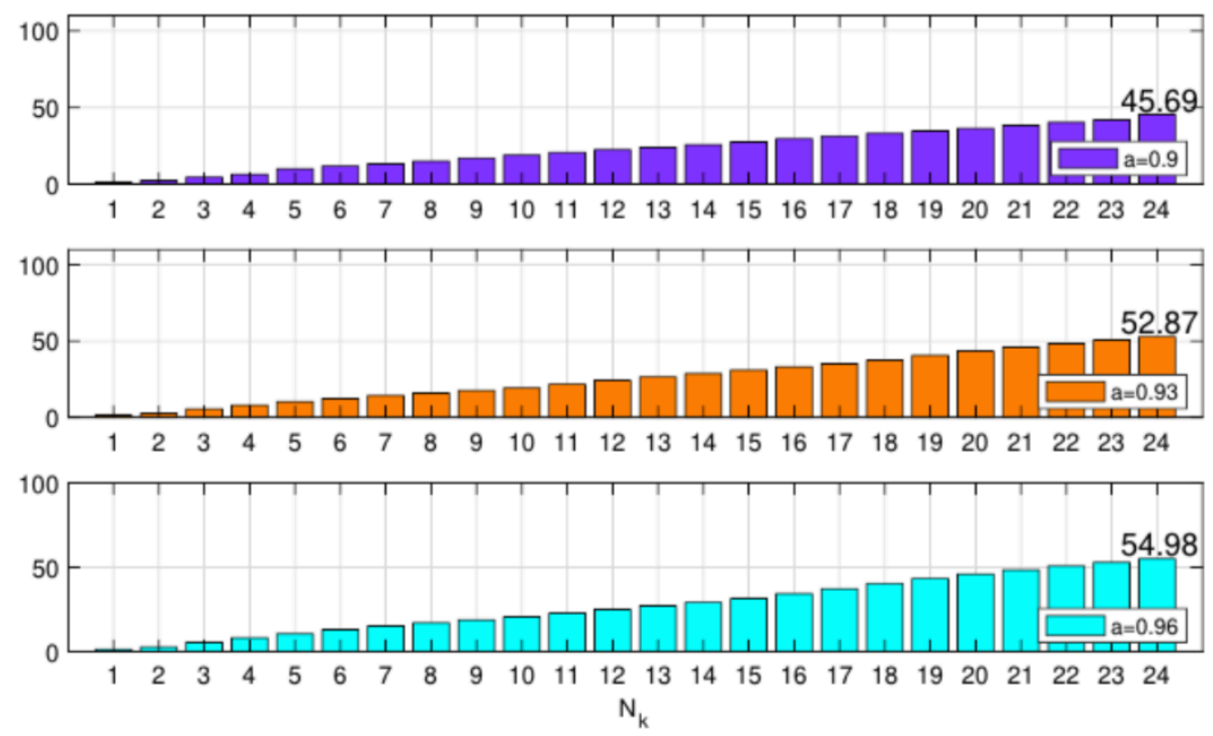

Figure 16. Scenario 1 cost function in $N_k = 24$

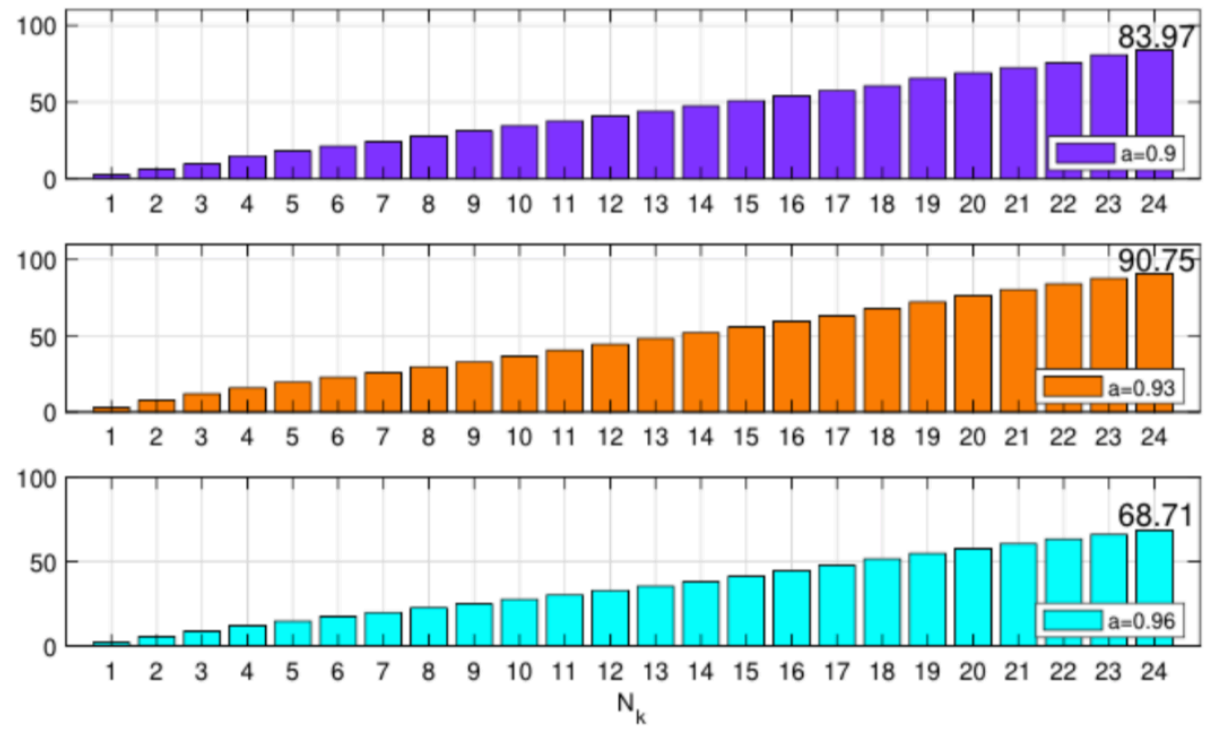

Figure 17. Scenario 2 cost function in $N_k = 24$

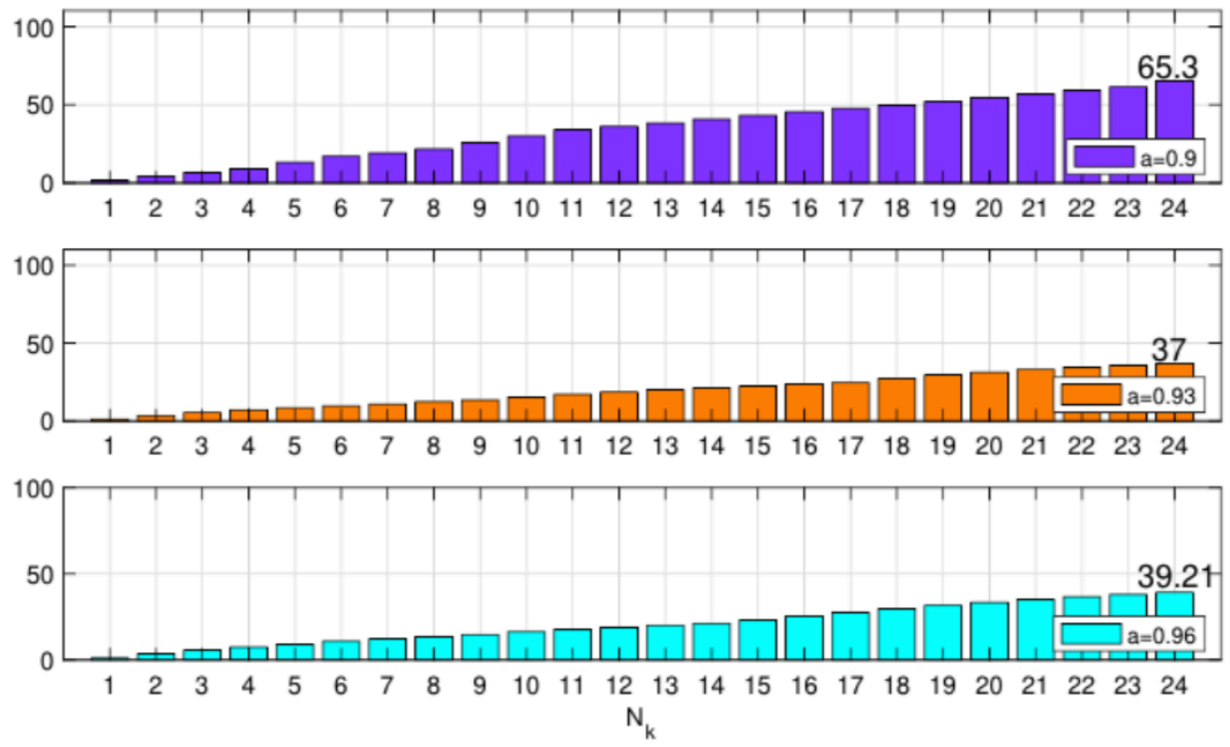

Figure 18. Scenario 3 cost function in $N_k = 24$

*5) Diesel Generators and Battery Energy Storage System*

To respond to customer's demand, there are DGs and BESSs to recompense power deficiency. The power of each DG and BESS are classified into bids. Here, DG bids are 50, 100, 150, 200 kW and BESS bids are 10, 20, 30, 40 kW. The choice of bids in Figure 19 are made considering their costs in equations (1) to (4).

Scenarios 2 and 3 are enabled to optimize the responded demand using parameter $J_3$ in total function; hence, as the RES supports the same amount of power in both scenarios 1 and 2, the system is forced to use the more capacity of DGs and BESSs in scenario 2 compared to scenario 1 as demonstrated in Figures 19(a) to 19(d). Also, in scenario 3, due to existence of uncertainties the RES performance decreases and the DGs and

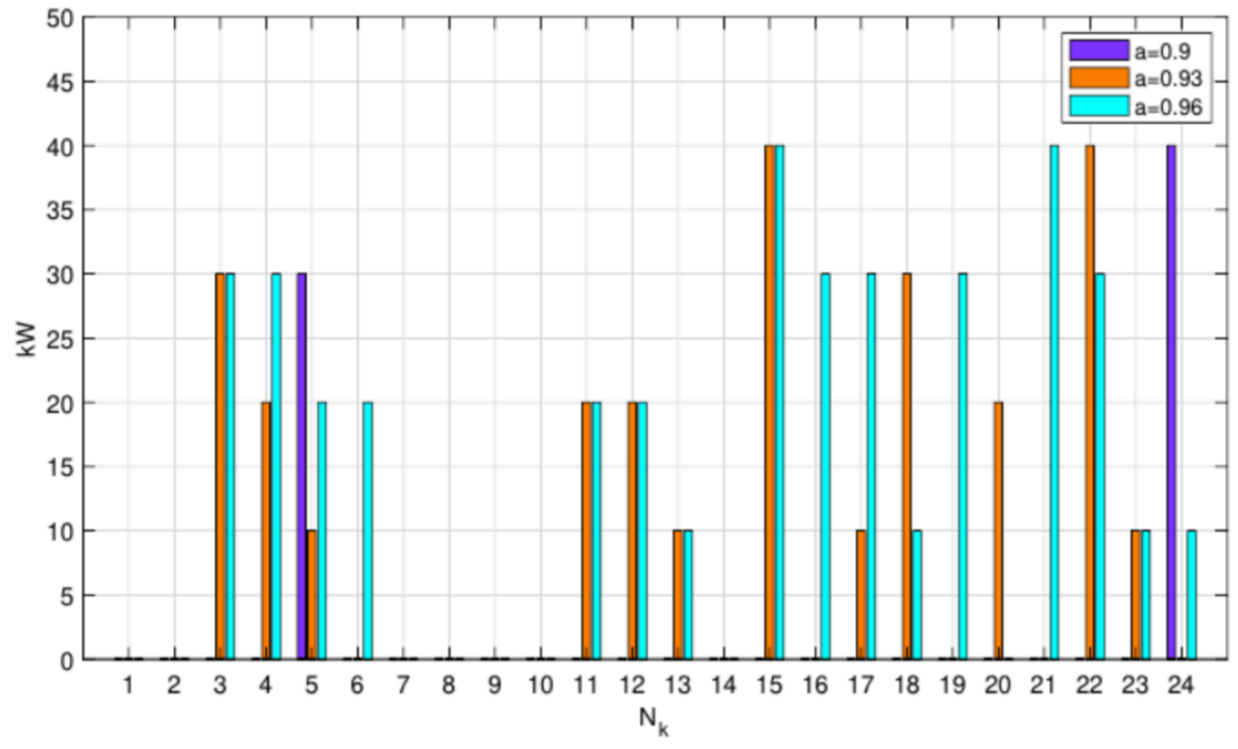

(a) BESS for scenario 1 in $N_k = 24$

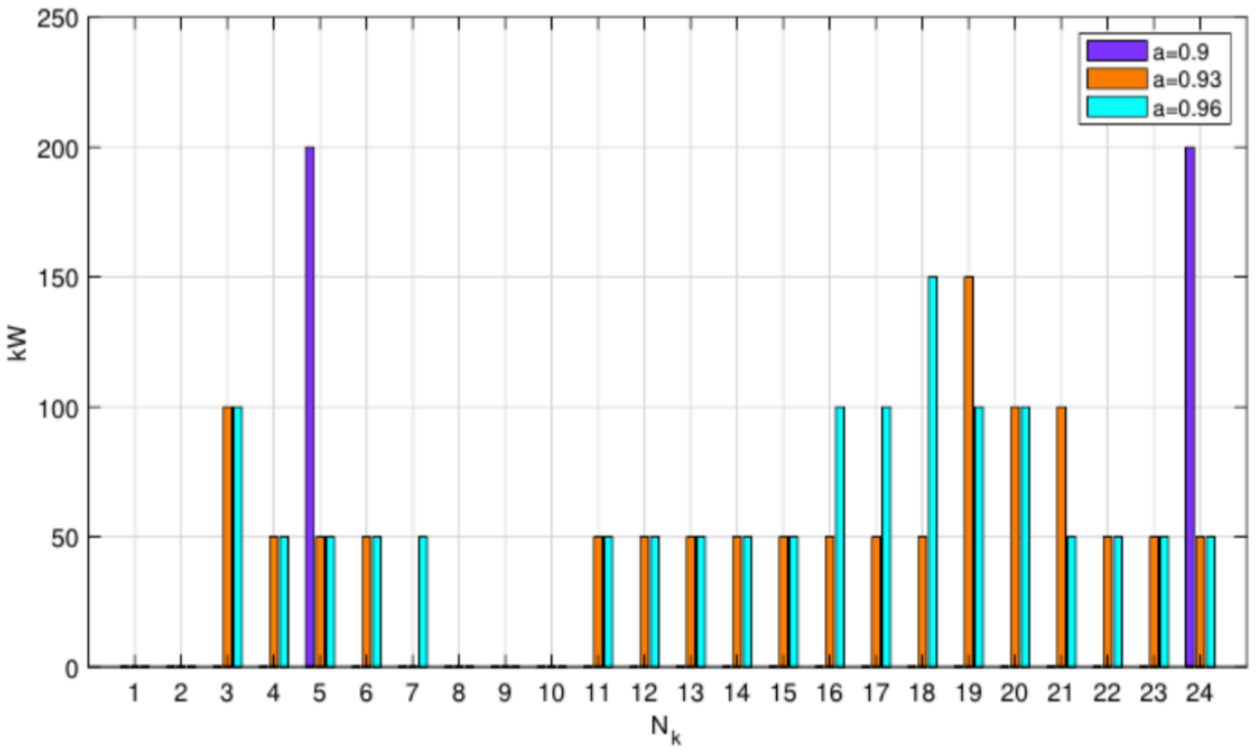

(b) DG for scenario 1 in $N_k = 24$

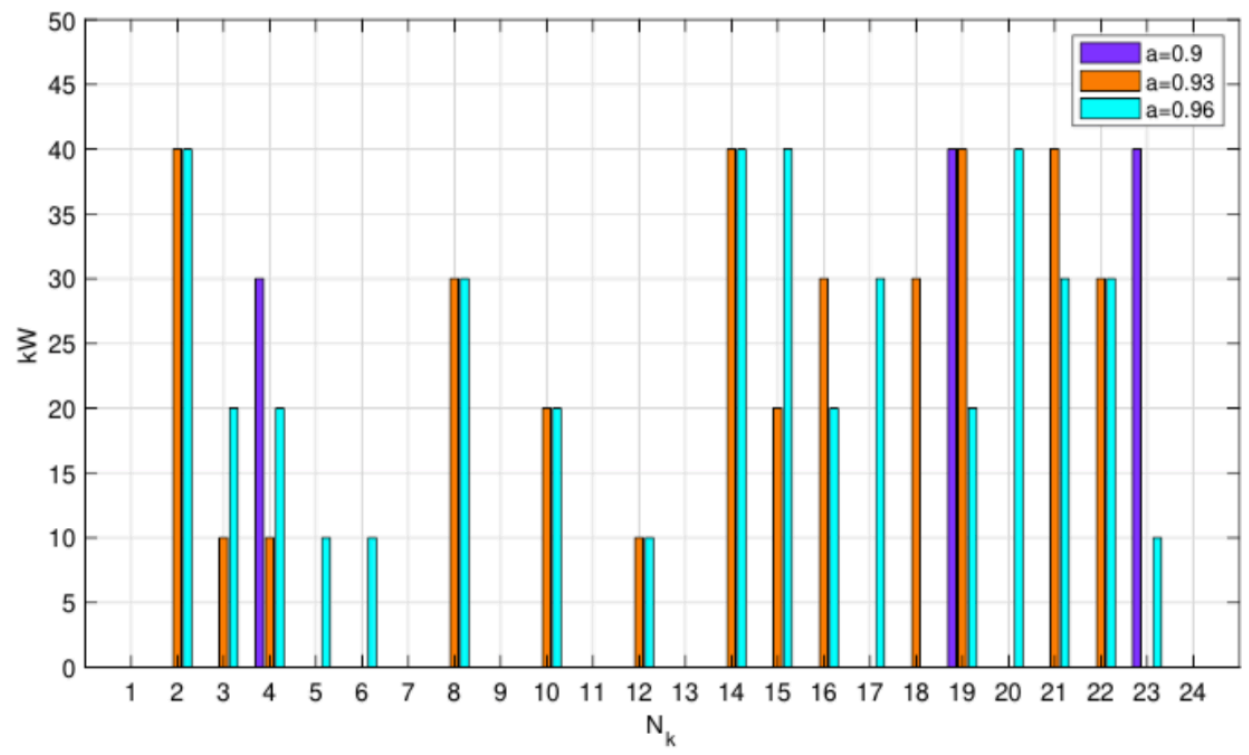

(c) BESS for scenario 2 in $N_k = 24$

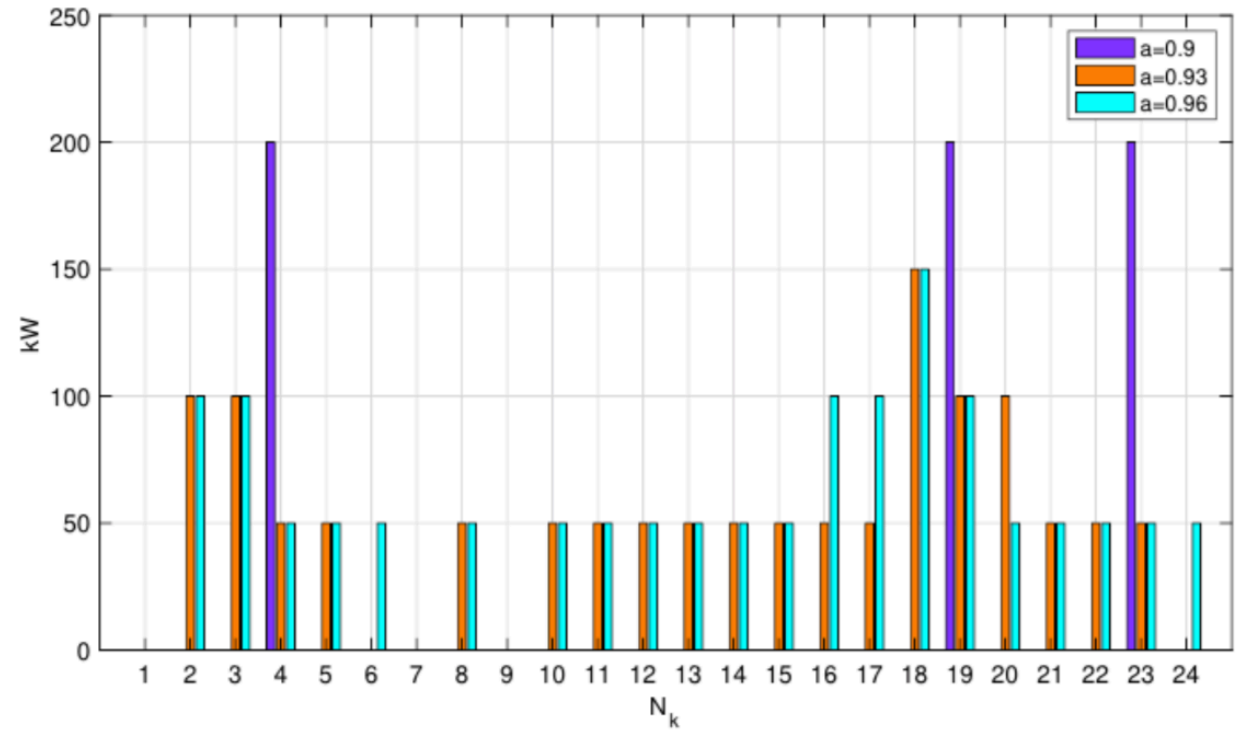

(d) DG for scenario 2 in $N_k = 24$

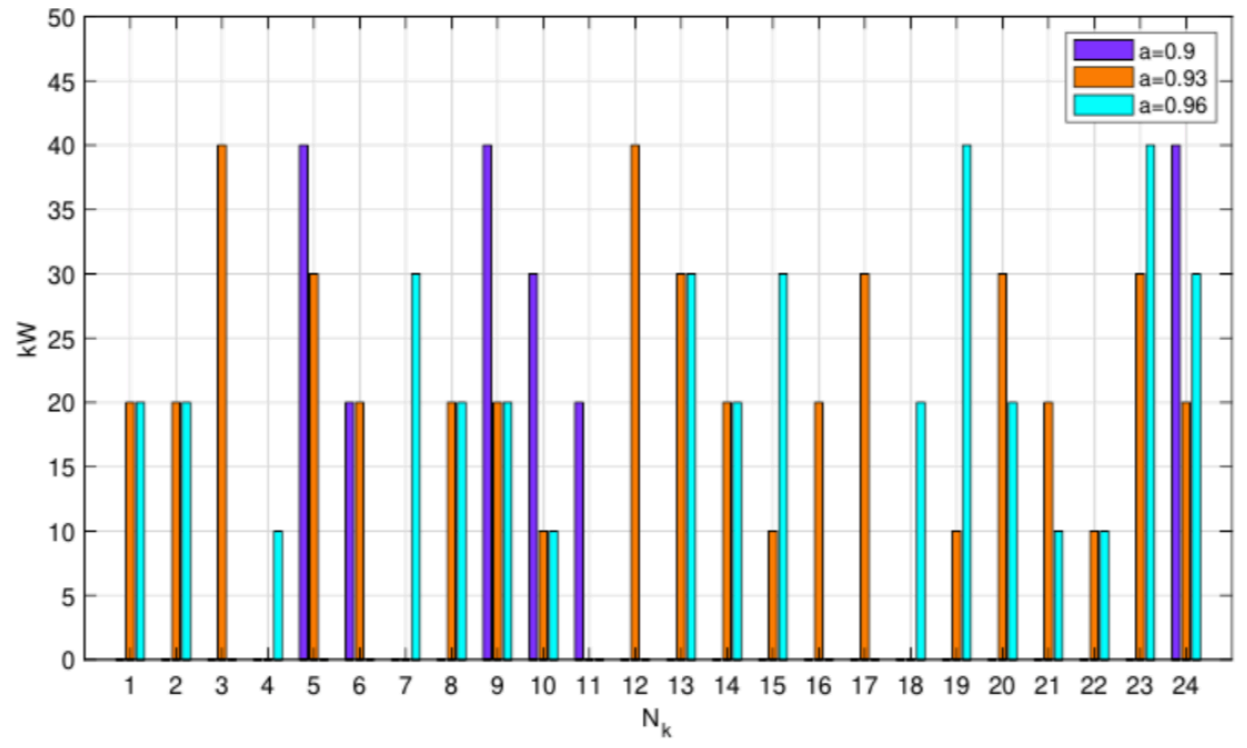

(e) BESS for scenario 3 in $N_k = 24$

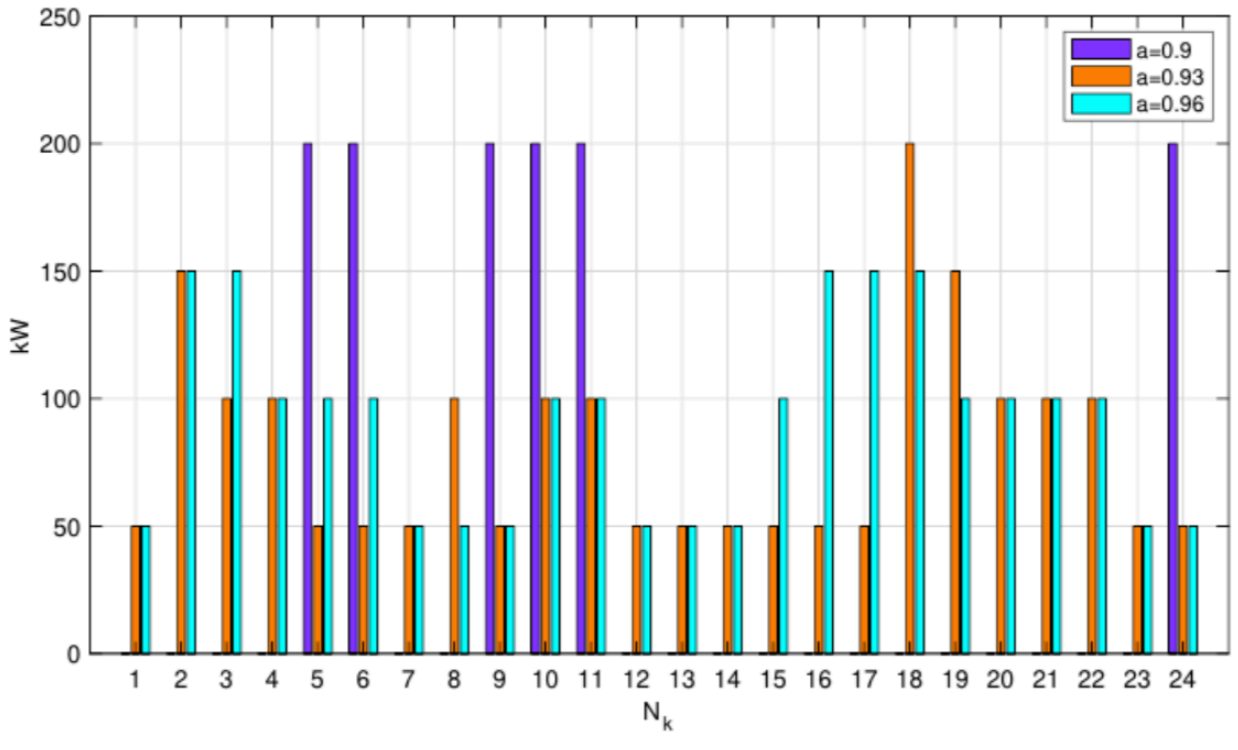

(f) DG for scenario 3 in $N_k = 24$

Figure 19. DG and BESS in different scenarios

| | RES | Uncrtn. | Market price | SOC | | | Cost function | | | DG\kW | | | BESS\kW | | |
|---|---|---|---|---|---|---|---|---|---|---|---|---|---|---|---|
| Scenario | MW | +\− | unconst./ const. | 0.9 | 0.93 | 0.96 | 0.9 | 0.93 | 0.96 | 0.9 | 0.93 | 0.96 | 0.9 | 0.93 | 0.96 |
| #1 | 4.27 | − | const. | 0.7527 | 0.7586 | 0.7737 | 45.69 | 52.87 | 54.99 | 16667 | 47917 | 54167 | 2917 | 10833 | 15833 |
| #2 | 4.27 | − | unconst. | 0.7607 | 0.7765 | 0.7918 | 83.97 | 90.75 | 68.71 | 25000 | 52083 | 58333 | 4533 | 14583 | 17500 |
| #3 | 2.94 | + | unconst. | 0.7607 | 0.7765 | 0.7918 | 65.3 | 37 | 39.21 | 50000 | 81250 | 91667 | 79167 | 18750 | 15833 |

Table 6. Scenarios' results

BESSs are scheduled to involve more as shown in Figures 19(e) and 19(f).

Table 6 illustrates the results in this part. In cases 2 and 3, the average SOC level is increased to admit that the price varying market is enabled to support a higher level of demand compared to the constant price market. In the RES column, the loss of power in case 3 is damaging and it results in more participation of DGs and BESSs as it is highlighted for $a = 0.93$.

### *6) Comparative results*

In this part, the energy management in three scenarios is compared for a class of DERs with $a = 0.9$. In case 1, the price market is constant; however, in other methods, it changes dynamically. The $p^{clear}$ for the next 24 hours is specified by predicting the maximum charge for DERs during this cycle. Figures 20 and 21 show the market price and customers price in both constant and inconstant cases.

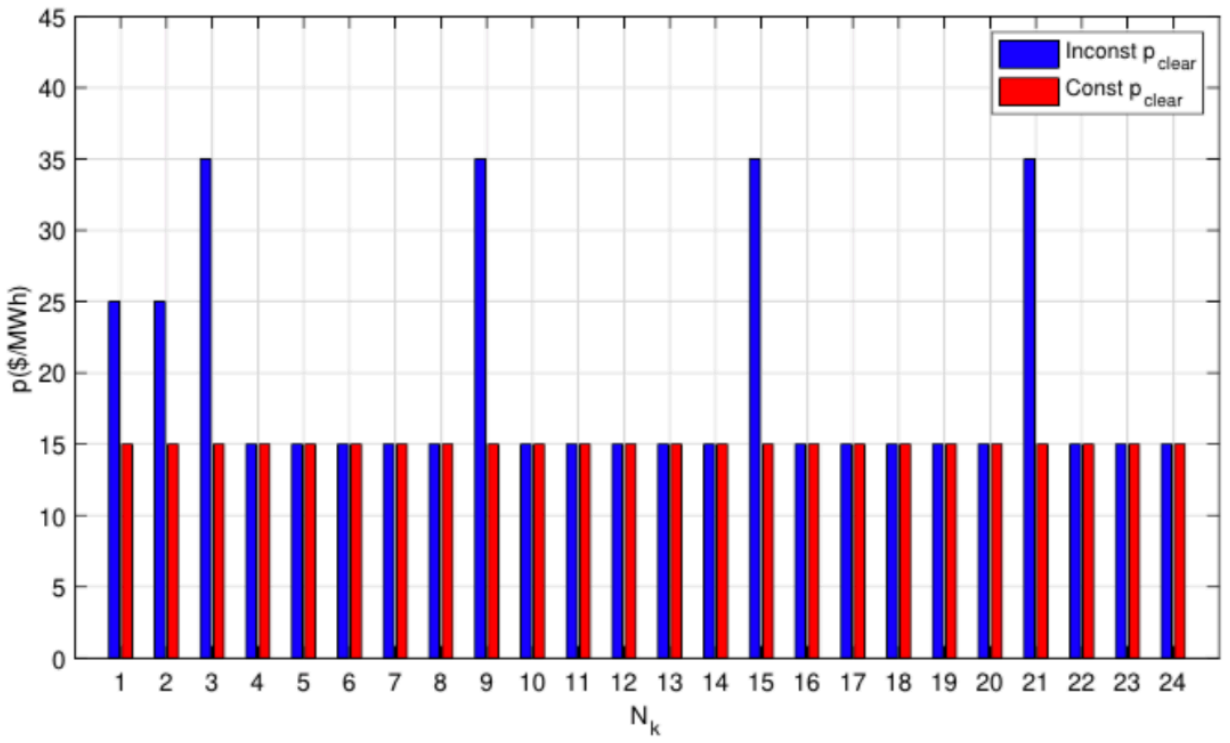

Figure 20. $p^{clear}$ in $N_k = 24$ for $a = 0.9$

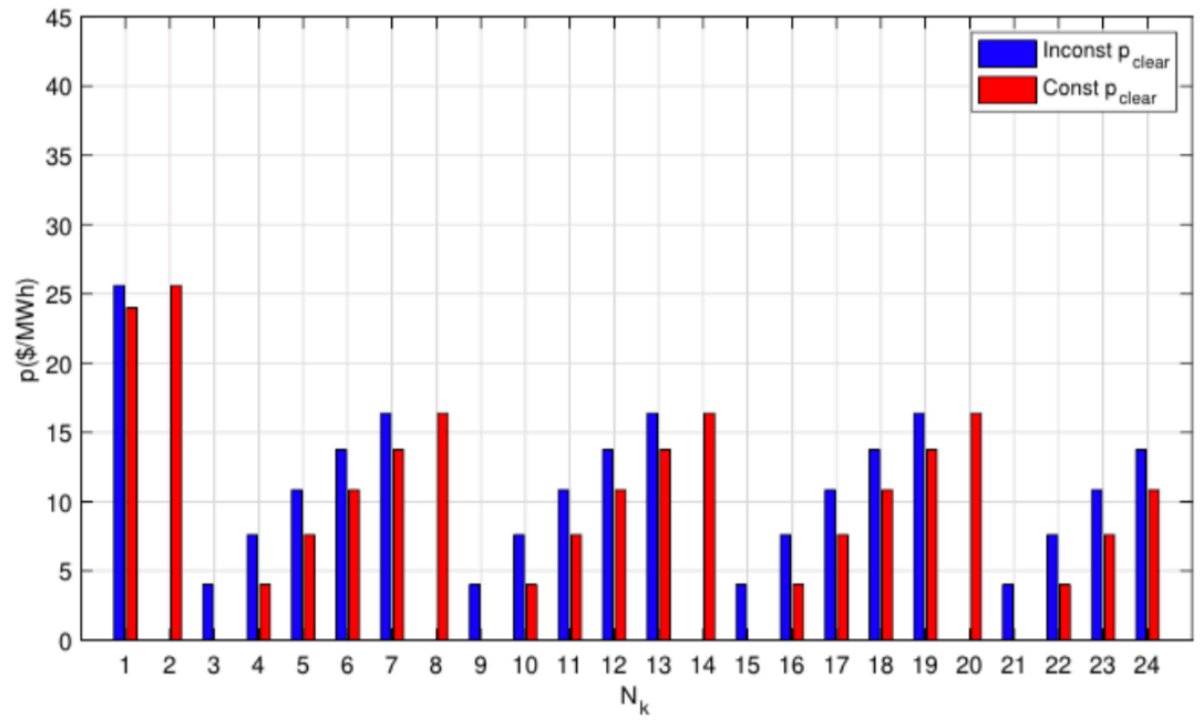

Figure 21. Price in $N_k = 24$ for $a = 0.9$

In case 1 as the market price is constant, there is no optimization on the state of charges to maximize them through $J_3$ in (4). Thus, the total value of cost function in the first case is lower than in other cases as shown in Figure 24. Dynamic market price enables us to predict maximum affordable demand and assign suitable bids as a clear price; hence, scenario 2 is able to support higher demand. As the power provided by RES is assumed to be constant in both cases 1 and 2, the DGs and BESSs have to take more responsibility as demonstrated in Figures 22 and 23. In addition to that, in scenario 3 the uncertainty of RES is included where they can not perform with their full potential. therefore, DGs and BESSs are even more included in this method compared to scenario 2.

It is noteworthy, that all scenarios are compelling and the differences are advantageous. Although the first scenario debilitates us to find the maximum economical price, it works much faster and has low computational cost than other scenarios as it is inessential to predict costs in the next 24 hours in each step for all provided market bids. The second scenario considers DERs state of charges as an important part of optimization. Apart from optimizing SOCs, this fact empowers customers to participate and pay less on electricity bills. Despite the higher volume of computational cost, the third scenario works sturdy under RES uncertainties where the other two methods are vulnerable.

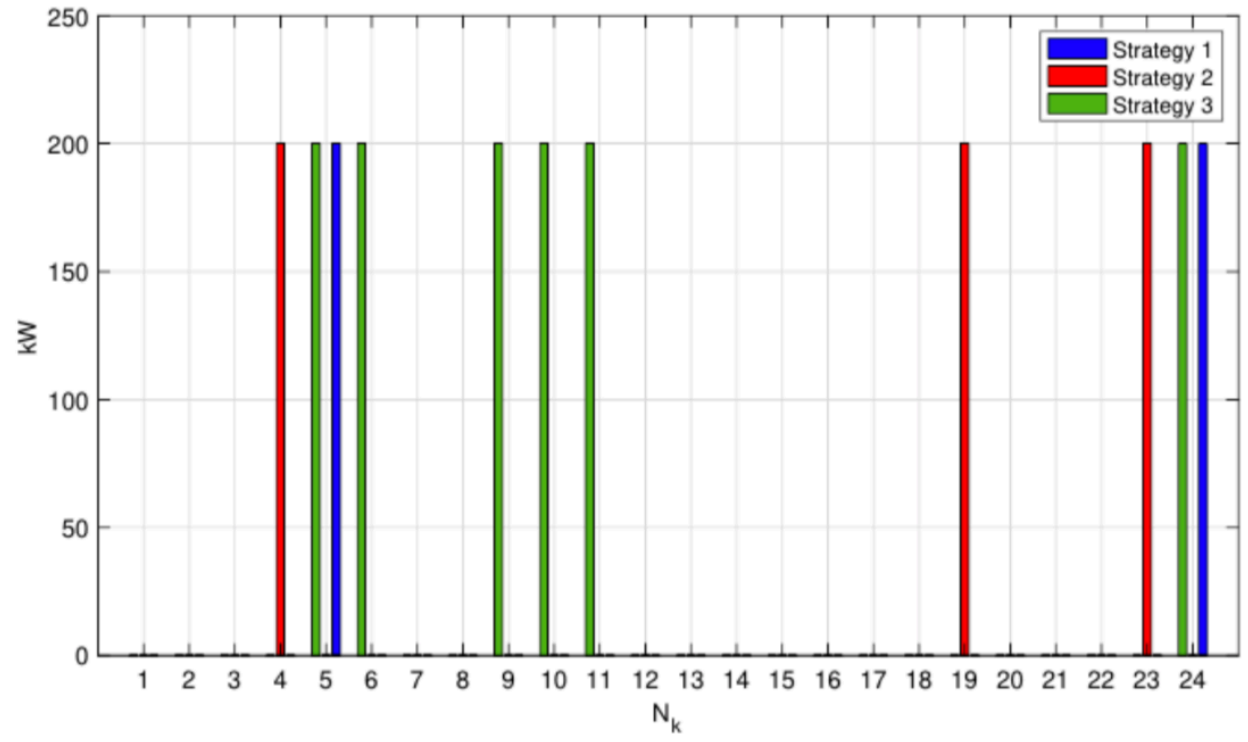

Figure 22. DG in $N_k = 24$ for $a = 0.9$

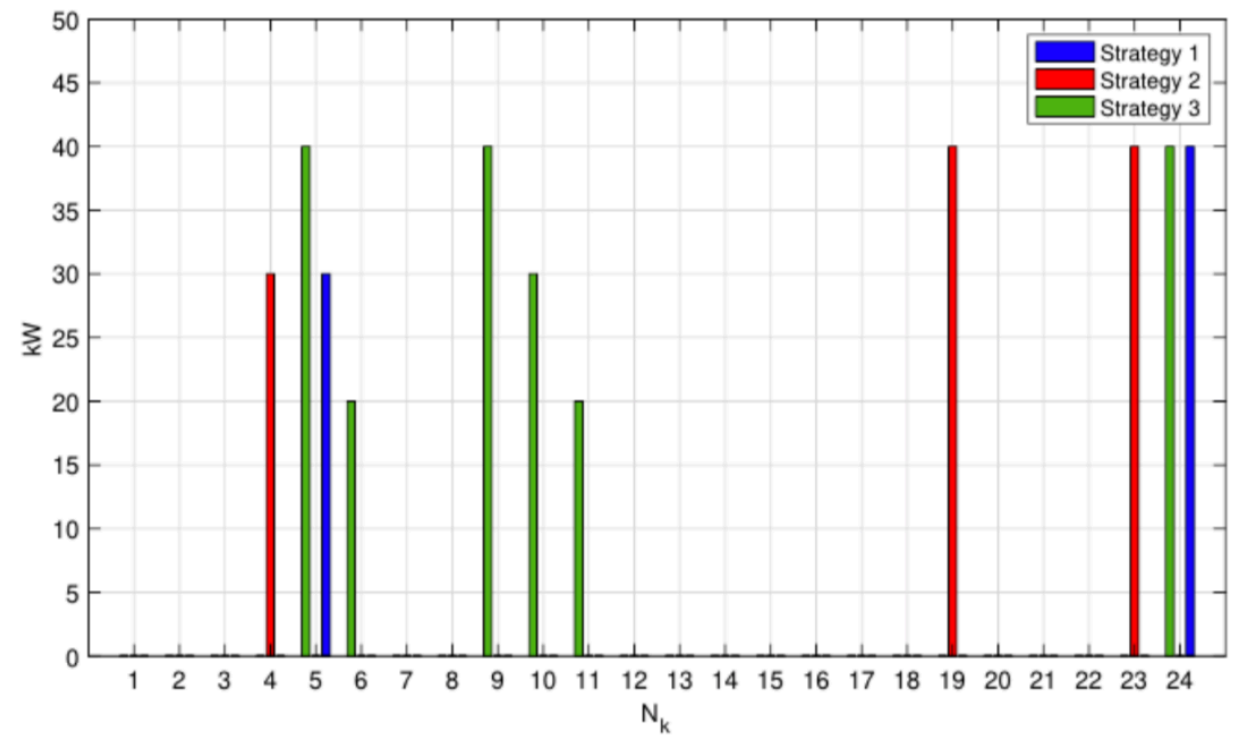

Figure 23. BESS in $N_k = 24$ for $a = 0.9$

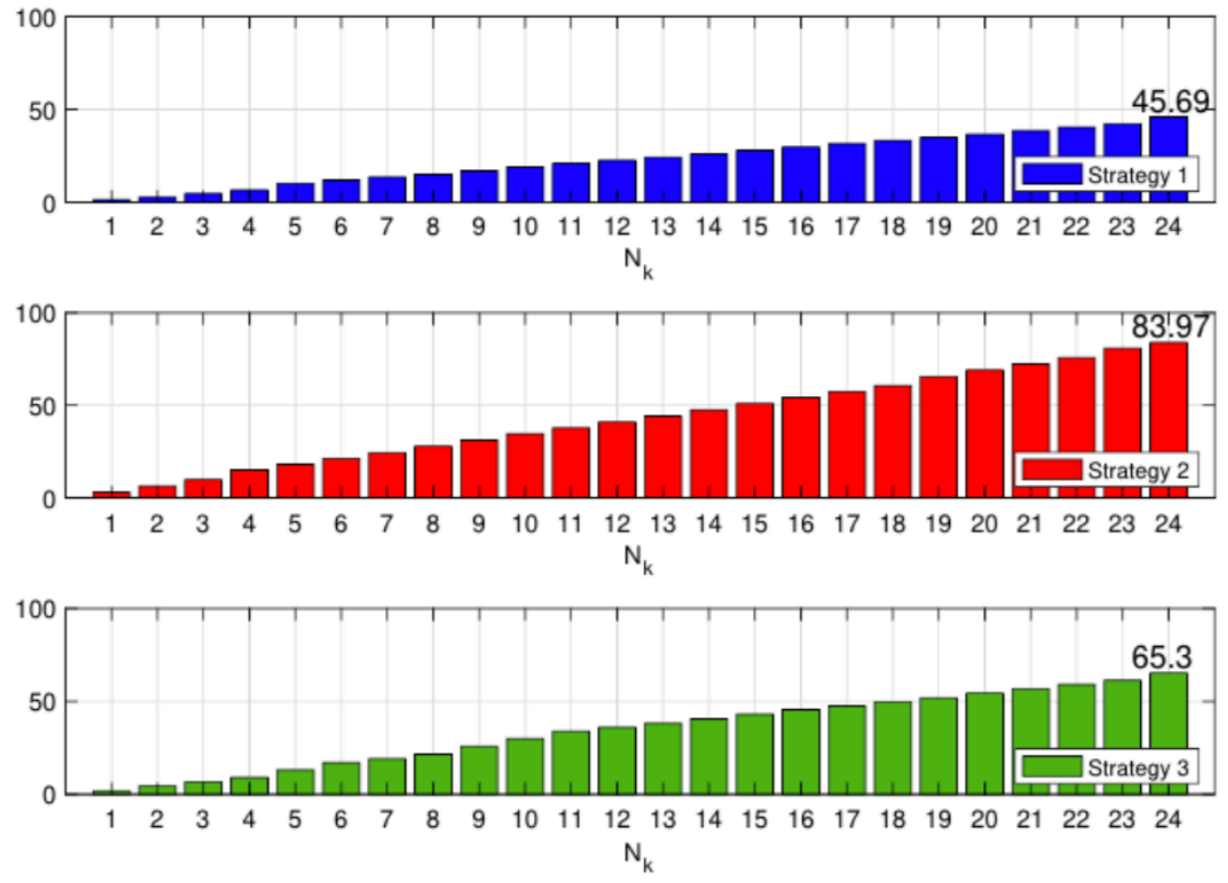

Figure 24. Cost function in $N_k = 24$ for $a = 0.9$

## V. CONCLUSION

This paper proposes a reliable energy scheduling framework to achieve an appropriate daily electricity consumption with maximum affordable demand response. Renewable and non-renewable energy resources are available to respond to customer's demands using a different class of methods to manage energy during the time. An optimal operation problem is investigated using MPC to determine which dispatchable unit should be operated at what time and at what power level while satisfying practical constraints. This research investigates the energy management of these systems in three complementary scenarios. The first and second scenarios are respectively suitable for a market with a constant and inconstant price. In addition to that, the third scenario is proposed to consider the role of uncertainties in RES and it is designed to recompense the power shortage using non-renewable resources. The validity of methods is explored in a residential area for 24 hours and the results thoroughly demonstrated the competence of the proposed approach for decreasing the operation cost.

## REFERENCES

[1] D. D'Agostino, B. Cuniberti, and P. Bertoldi, "Energy consumption and efficiency technology measures in european non-residential buildings," *Energy and Buildings*, vol. 153, pp. 72–86, 2017.

[2] R. Godina, E. M. Rodrigues, E. Pouresmaeil, J. C. Matias, and J. P. Catalão, "Model predictive control home energy management and optimization strategy with demand response," *Applied Sciences*, vol. 8, no. 3, p. 408, 2018.

[3] C. Cecati, C. Citro, and P. Siano, "Combined operations of renewable energy systems and responsive demand in a smart grid," *IEEE Transactions on sustainable energy*, vol. 2, no. 4, pp. 468–476, 2011.

[4] A. Chaouachi, R. M. Kamel, R. Andoulsi, and K. Nagasaka, "Multiobjective intelligent energy management for a microgrid," *IEEE Transactions on Industrial Electronics*, vol. 60, no. 4, pp. 1688–1699, 2012.

[5] R. Palma-Behnke, C. Benavides, F. Lanas, B. Severino, L. Reyes, J. Llanos, and D. Sáez, "A microgrid energy management system based on the rolling horizon strategy," *IEEE Transactions on smart grid*, vol. 4, no. 2, pp. 996–1006, 2013.

[6] A. Khodaei, "Microgrid optimal scheduling with multi-period islanding constraints," *IEEE Transactions on Power Systems*, vol. 29, no. 3, pp. 1383–1392, 2013.

[7] W. Shi, X. Xie, C.-C. Chu, and R. Gadh, "Distributed optimal energy management in microgrids," *IEEE Transactions on Smart Grid*, vol. 6, no. 3, pp. 1137–1146, 2014.

[8] W. Shi, N. Li, C.-C. Chu, and R. Gadh, "Real-time energy management in microgrids," *IEEE Transactions on Smart Grid*, vol. 8, no. 1, pp. 228–238, 2015.

[9] M. Amini, J. Frye, M. D. Ilic, and O. Karabasoglu, "Smart residential energy scheduling utilizing two stage mixed integer linear programming," in *2015 North American Power Symposium (NAPS)*. IEEE, 2015, pp. 1–6.

[10] Z. Wang, B. Chen, J. Wang, M. M. Begovic, and C. Chen, "Coordinated energy management of networked microgrids in distribution systems," *IEEE Transactions on Smart Grid*, vol. 6, no. 1, pp. 45–53, 2014.

[11] W. Su, J. Wang, and J. Roh, "Stochastic energy scheduling in microgrids with intermittent renewable energy resources," *IEEE Transactions on Smart grid*, vol. 5, no. 4, pp. 1876–1883, 2013.

[12] Y. Zhang, N. Gatsis, and G. B. Giannakis, "Robust energy management for microgrids with high-penetration renewables," *IEEE transactions on sustainable energy,* vol. 4, no. 4, pp. 944–953, 2013.

[13] REN21, "Renewables 2019: Global status report," *REN21 Secretariat*, 2019.

[14] N. Gatsis and G. B. Giannakis, "Residential load control: Distributed scheduling and convergence with lost AMI messages," *IEEE Transactions on Smart Grid,* vol. 3, no. 2, pp. 770–786, 2012.

[15] C. Chen, J. Wang, Y. Heo, and S. Kishore, "MPC-based appliance scheduling for residential building energy management controller," *IEEE Transactions on Smart Grid,* vol. 4, no. 3, pp. 1401–1410, 2013.

[16] T.-H. Chang, M. Alizadeh, and A. Scaglione, "Real-time power balancing via decentralized coordinated home energy scheduling," *IEEE Transactions on Smart Grid*, vol. 4, no. 3, pp. 1490–1504, 2013.

[17] M. Rahmani-Andebili, "Scheduling deferrable appliances and energy resources of a smart home applying multi-time scale stochastic model predictive control," *Sustainable Cities and Society*, vol. 32, pp. 338–347, 2017.

[18] P. Samadi, H. Mohsenian-Rad, V. W. Wong, and R. Schober, "Tackling the load uncertainty challenges for energy consumption scheduling in smart grid," *IEEE Transactions on Smart Grid*, vol. 4, no. 2, pp. 1007–1016, 2013.

[19] M. Ding, Y. Zhang, M. Mao, X. Liu, and N. Xu, "Economic operation optimization for microgrids including Na/S battery storage," *Proceedings of the CSEE*, vol. 31, no. 4, pp. 7–14, 2011.

[20] Y. Xiang, J. Liu, and Y. Liu, "Robust energy management of microgrid with uncertain renewable generation and load," *IEEE Transactions on Smart Grid*, vol. 7, no. 2, pp. 1034–1043, 2015.

[21] A. Parisio, C. Del Vecchio, and A. Vaccaro, "A robust optimization approach to energy hub management," *International Journal of Electrical Power & Energy Systems*, vol. 42, no. 1, pp. 98–104, 2012.

[22] R. Wang, P. Wang, and G. Xiao, "A robust optimization approach for energy generation scheduling in microgrids," *Energy Conversion and Management*, vol. 106, pp. 597–607, 2015.

[23] Y. Wang, Q. Xia, and C. Kang, "Unit commitment with volatile node injections by using interval optimization," *IEEE Transactions on Power Systems*, vol. 26, no. 3, pp. 1705–1713, 2011.

[24] N. Growe-Kuska, H. Heitsch, and W. Romisch, "Scenario reduction and scenario tree construction for power management problems," in *2003 IEEE Bologna Power Tech Conference Proceedings*, vol. 3. IEEE, 2003, pp. 7–pp.

[25] H. Gao, J. Liu, Z. Wei, Y. Cao, W. Wang, and S. Huang, "A security-constrained dispatching model for wind generation units based on extreme scenario set optimization," *Power System Technology*, vol. 6, p. 018, 2013.

[26] J. M. Morales, S. Pineda, A. J. Conejo, and M. Carrion, "Scenario reduction for futures market trading in electricity markets," *IEEE Transactions on Power Systems*, vol. 24, no. 2, pp. 878–888, 2009.

[27] B. Venkatesh, P. Yu, H. Gooi, and D. Choling, "Fuzzy MILP unit commitment incorporating wind generators," *IEEE transactions on power systems*, vol. 23, no. 4, pp. 1738–1746, 2008.

[28] R. H. Liang and J. H. Liao, "A fuzzy-optimization approach for generation scheduling with wind and solar energy systems," *IEEE Transactions on Power Systems*, vol. 22, no. 4, pp. 1665–1674, 2007.

[29] T. Niknam, R. Azizipanah-Abarghooee, and M. R. Narimani, "An efficient scenario-based stochastic programming framework for multi-objective optimal micro-grid operation," *Applied Energy*, vol. 99, pp. 455–470, 2012.

[30] L. Fiorini and M. Aiello, "Energy management for user's thermal and power needs: A survey," *Energy Reports*, vol. 5, pp. 1048–1076, 2019.

[31] G. Cau, D. Cocco, M. Petrollese, S. K. Kær, and C. Milan, "Energy management strategy based on short-term generation scheduling for a renewable microgrid using a hydrogen storage system," *Energy Conversion and Management,* vol. 87, pp. 820–831, 2014.

[32] M. Meiqin, J. Meihong, D. Wei, and L. Chang, "Multi-objective economic dispatch model for a microgrid considering reliability," in *The 2nd International Symposium on Power Electronics for Distributed Generation Systems*. IEEE, 2010, pp. 993–998.

[33] F. Brahman, M. Honarmand, and S. Jadid, "Optimal electrical and thermal energy management of a residential energy hub, integrating demand response and energy storage system," *Energy and Buildings*, vol. 90, pp. 65–75, 2015.

[34] A. Anvari-Moghaddam, A. Rahimi-Kian, M. S. Mirian, and J. M. Guerrero, "A multi-agent based energy management solution for integrated buildings and microgrid system," *Applied energy*, vol. 203, pp. 41–56, 2017.

[35] E. Shirazi and S. Jadid, "Optimal residential appliance scheduling under dynamic pricing scheme via HEMDAS," *Energy and Buildings*, vol. 93, pp. 40–49, 2015.

[36] I. Prodan and E. Zio, "A model predictive control framework for reliable microgrid energy management," *International Journal of Electrical Power & Energy Systems,* vol. 61, pp. 399–409, 2014.

[37] S. M. Hosseini, R. Carli, and M. Dotoli, "Model predictive control for real-time residential energy scheduling under uncertainties," in *2018 IEEE International Conference on Systems, Man, and Cybernetics (SMC).* IEEE, 2018, pp. 1386–1391.

[38] A. Hooshmand, H. A. Malki, and J. Mohammadpour, "Power flow management of microgrid networks using model predictive control," *Computers & Mathematics with Applications*, vol. 64, no. 5, pp. 869– 876, 2012.

[39] A. Moradmand, A. Ramezani, H. S. Nezhad, and A. Sardashti, "Fault tolerant Kalman filter-based distributed predictive control in power systems under governor malfunction," in *2019 6th International Conference on Control, Instrumentation and Automation (ICCIA).* IEEE, 2019, pp. 1–6.

[40] H. S. Nezhad, A. Sardashti, A. Ramezani, and A. Moradmand, "Sensor fault detection in a distributed power system via decentralized Kalman filters," in *2019 6th International Conference on Control, Instrumentation and Automation (ICCIA).* IEEE, 2019, pp. 1–6.

[41] A. Sardashti, A. Ramezani, H. S. Nezhad, and A. Moradmand, "Observer-based sensor fault detection in islanded AC microgrids using online recursive estimation," in *2019 6th International Conference on Control, Instrumentation and Automation (ICCIA).* IEEE, 2019, pp. 1–6.

[42] H. Hao, B. M. Sanandaji, K. Poolla, and T. L. Vincent, "A generalized battery model of a collection of thermostatically controlled loads for providing ancillary service," in *2013 51st Annual Allerton Conference on Communication, Control, and Computing (Allerton).* IEEE, 2013, pp. 551–558.

[43] M. S. Nazir and I. A. Hiskens, "A dynamical systems approach to modeling and analysis of transactive energy coordination," *IEEE Transactions on Power Systems,* vol. 34, no. 5, pp. 4060–4070, 2018.

[44] J. L. Mathieu, M. Kamgarpour, J. Lygeros, G. Andersson, and D. S.Callaway, "Arbitraging intraday wholesale energy market prices with aggregations of thermostatic loads," *IEEE Transactions on Power Systems,* vol. 30, no. 2, pp. 763–772, 2014.

[45] D. Sera, R. Teodorescu, and P. Rodriguez, "PV panel model based on datasheet values," in *2007 IEEE international symposium on industrial electronics.* IEEE, 2007, pp. 2392–2396.

[46] W. T. P. Calculations, "RWE npower renewables," *Mechanical and Electrical Engineering Power Industry, The Royal Academy of Engineering,* 2012.